\documentclass[%
 superscriptaddress,
 reprint,
 floatfix,
 amsmath,amssymb,
 aip,
 citeautoscript,
 article
]{revtex4-1}

\usepackage{graphicx}
\usepackage{dcolumn}
\usepackage{bm}
\usepackage[x11names]{xcolor}

\begin{document}

\preprint{}

\title{Dynamical Bonding Driving Mixed Valency in a Metal Boride}

\author{Paul J. Robinson}
    \thanks{Current Address: Department of Chemistry, Columbia University, New York, New York 10027, USA}
    \affiliation{Department of Chemistry and Biochemistry, University of California Los Angeles, Los Angeles, California 90095, USA}
\author{Julen Munarriz}
    \affiliation{Department of Chemistry and Biochemistry, University of California Los Angeles, Los Angeles, California 90095, USA}
\author{Michael E. Valentine}
    \affiliation{Institute for Quantum Matter, Department of Physics and Astronomy, The Johns Hopkins University, Baltimore, Maryland 21218, USA}
\author{Austin Granmoe} 
    \affiliation{Institute for Quantum Matter, Department of Physics and Astronomy, The Johns Hopkins University, Baltimore, Maryland 21218, USA}
\author{Natalia Drichko}
    \email{drichko@jhu.edu, mcqueen@jhu.edu, ana@chem.ucla.edu }
    \affiliation{Institute for Quantum Matter, Department of Physics and Astronomy, The Johns Hopkins University, Baltimore, Maryland 21218, USA}
\author{Juan R. Chamorro}
    \affiliation{Institute for Quantum Matter, Department of Physics and Astronomy, The Johns Hopkins University, Baltimore, Maryland 21218, USA}
    \affiliation{Department of Chemistry, The Johns Hopkins University, Baltimore, Maryland 21218, USA}
\author{Priscila F. Rosa}
    \affiliation{Los Alamos National Laboratory, Los Alamos, NM 87545, USA}
\author{Tyrel M. McQueen}
    \email{drichko@jhu.edu, mcqueen@jhu.edu, ana@chem.ucla.edu }
    \affiliation{Institute for Quantum Matter, Department of Physics and Astronomy, The Johns Hopkins University, Baltimore, Maryland 21218, USA}
    \affiliation{Department of Chemistry, The Johns Hopkins University, Baltimore, Maryland 21218, USA}
    \affiliation{Department of Materials Science and Engineering, The Johns Hopkins University, Baltimore, Maryland 21218, USA}
\author{Anastassia N. Alexandrova}
    \email{drichko@jhu.edu, mcqueen@jhu.edu, ana@chem.ucla.edu }
    \affiliation{Department of Chemistry and Biochemistry, University of California Los Angeles, Los Angeles, California 90095, USA}
    \affiliation{California NanoSystems Institute, Los Angeles, California 90095, USA}

\date{\today}

\begin{abstract}
Samarium hexaboride is an anomaly, having many exotic and seemingly mutually incompatible properties. It was proposed to be a mixed-valent semiconductor, and later a topological Kondo insulator, and yet has a Fermi surface despite being an insulator. We propose a new and unified understanding of SmB$_6$ centered on the hitherto unrecognized dynamical bonding effect: the coexistence of two Sm-B bonding modes within $\text{SmB}_6$, corresponding to different oxidation states of the Sm. The mixed valency arises in SmB$_6$ from thermal population of these distinct minima enabled by motion of B. Our model simultaneously explains the thermal valence fluctuations, appearance of magnetic Fermi surface, excess entropy at low temperatures, pressure-induced phase transitions, and related features in Raman spectra and their unexpected dependence on temperature and boron isotope.  
\end{abstract}
\maketitle

\section*{Introduction}
Samarium hexaboride ($\text{SmB}_6$) is a famously confusing solid, having attracted constant attention since its characterization \cite{1nickerson1971physical}. Nearly fifty years later, the ground state of $\text{SmB}_6$ remains unresolved \cite{2cooley1995sm,3martin1979theory}. $\text{SmB}_6$ as a Kondo insulator is an intuitive thought: a three-band model predicts that the localized  \textit{f}-orbital peak at the Fermi level will hybridize with the \textit{d}-orbitals and open a gap, making an insulator. This would normally preclude the existence of a Fermi surface and thus remove any quantum oscillations, a signature of Landau quantization of a Fermi surface in a magnetic field. However, $\text{SmB}_6$ displays quantum oscillations \cite{4li2014two,5jiang2013observation,6stern2017surface,7kim2014topological,8park2016topological}.
While a 2D Fermi surface can be a result of the presence of topological surface $\text{states,}^{7}$ a 3D Fermi surface \cite{9tan2015unconventional,10hartstein2018fermi} is in contradiction with apparently insulating properties of the material.  Many theories point to a topological type Kondo insulator allowing conduction only on the surface states, and some also point to composite particle excitons\cite{11knolle2017excitons}, 
and even call $\text{SmB}_6$ a Majorana Fermion sea.\cite{12baskaran2015majorana,13erten2017skyrme} Other phenomena in $\text{SmB}_6$ are a resistivity plateau\cite{14dzero2016topological}, and an anomalous peak in the specific heat\cite{15phelan2014correlation}, at T $\sim$ 20 K.

The central enigmatic property of $\text{SmB}_6$ is its homogeneous mixed valency. The lattice of $\text{SmB}_6$ is cubic, but spectroscopic studies reveal two distinct oxidation states, $\text{Sm}^{2+}$ and $\text{Sm}^{3+}$ without an intermediate $\text{Sm}^{2.5+}$ state. In contrast to other mixed-valence compounds, $\text{SmB}_6$ does not show charge order. The ratio between the two valencies responds to T and p \cite{16mizumaki2009temperature,17butch2016pressure}.
The average valency increases rapidly from around +2.5 at 0 K to a plateau at +2.6 above 100 K. This raises a chemical question: what type of mixed valency is the solid? The traditional Robin-Day scheme defines Types I-III mixed valency, corresponding to weak, medium, and strong couplings between two electronic states 
(e.g. $\text{M}^{2+}\text{M}^{3+}$ and $\text{M}^{3+}\text{M}^{2+}$), respectively \cite{18parthey2014quantum}.
However, the T-dependence puts in question how much the mixed valency in $\text{SmB}_6$ conforms to these known schemes. For some rare-earth and actinide compounds, valence instabilities and transitions with T, p, and composition, are documented \cite{19robinson1979valence},
and described mainly through band models, which consider valence transitions between a localized \textit{f}-state and a delocalized band state (neither being precisely defined) \cite{20ramirez1970metal}.
The energy of the localized states can rise in response to lattice compression, producing pressure effects. A seemingly disconnected theory for $\text{SmB}_6$ states that the ratio between $\text{Sm}^{2+}$ and $\text{Sm}^{3+}$ is held constant by the stiff, non-interacting boron network pressing against the hard sphere metal ions, while mixing $\text{Sm}^{2+}$ and $\text{Sm}^{3+}$---cations of different radii---minimized the lattice energy \cite{1nickerson1971physical}. The major assumption underlying this description is the lack of electronic Sm-B interactions. However, in other borides M-B interactions were found to be prominent\cite{21levine2009advancements,22gilman2006design,23brazhkin2002harder}, and sensitive to mechanical stress\cite{24robinson2017mystery,25lei2018understanding}. Hence, there has never been a satisfying microscopic theory explaining the origin of the mixed valency in $\text{SmB}_6$.

In this paper, we present a unifying dynamic bonding model of $\text{SmB}_6$, which readily explains its mixed valency, valence fluctuations and changes with T and p, and predicts many seemingly disparate properties such as the peak in the heat capacity, and the Raman scattering spectra associated with valency changes. We will build the solid from the ground up, starting with molecular clusters that reveal the key bonding elements present also in the solid, as suggested by Hoffmann, Pauling, and several followers\cite{26pauling1948metallic,27pauling1949resonating,28hoffmann1987chemistry,29BurdettChemicalBonding,30cox1987electronic,31canadellOrbtial}. Then, we reconstruct the chemically informed model of the full $\text{SmB}_6$ solid. The approach is necessitated by the intractably complex electronic structure of $\text{SmB}_6$: strongly multireference, and relativistic \cite{32jiao2016additional}. These effects inevitably present a problem for DFT, requiring the use of high level multi-reference and relativistic methods. Nonetheless, the latter are restricted to a few atoms, thus not being applicable to periodic systems such as $\text{SmB}_6$. Fortunately, we were able to construct a minimal model that captures the essential interactions in the real solid, for which the previous methodologies are affordable. The key to our model is the unique and dynamic bonding interactions between Sm and B\cite{33robinson2017smb6}.

\section*{Results and Discussion}
First, we examine the structural motifs present in $\text{SmB}_6$ and identify the key bonding interactions. A minimal cluster needn't be stoichiometrically identical to the solid; in fact the $\text{SmB}_6^-$ cluster has been studied theoretically and spectroscopically, and showed no structural resemblance with the material \cite{33robinson2017smb6}. $\text{SmB}_6$ has a cubic unit cell with a $\text{B}_6$ octahedron in the center; however, the B-B distance between cells is shorter than within the octahedron, suggestive of $\text{B}_2$ dimers. A plane-wave DFT calculation on this geometry reveals that the electrons are localized in-between the cells much more than in the octahedron (Fig. S1). Thus, considering $\text{SmB}_6$ as built of Sm and $\text{B}_2$ units brings us to $\text{SmB}_2^+$ as a minimal cluster model. Notice that the positive charge mitigates the undercoordinated environment of a gas phase cluster compared to the crystal. This cluster is small enough to explicitly calculate many-body and relativistic effects, and to qualitatively inform about the bonding motifs possible in the solid.

\begin{figure}
\centering 
\includegraphics[width=0.8\linewidth]{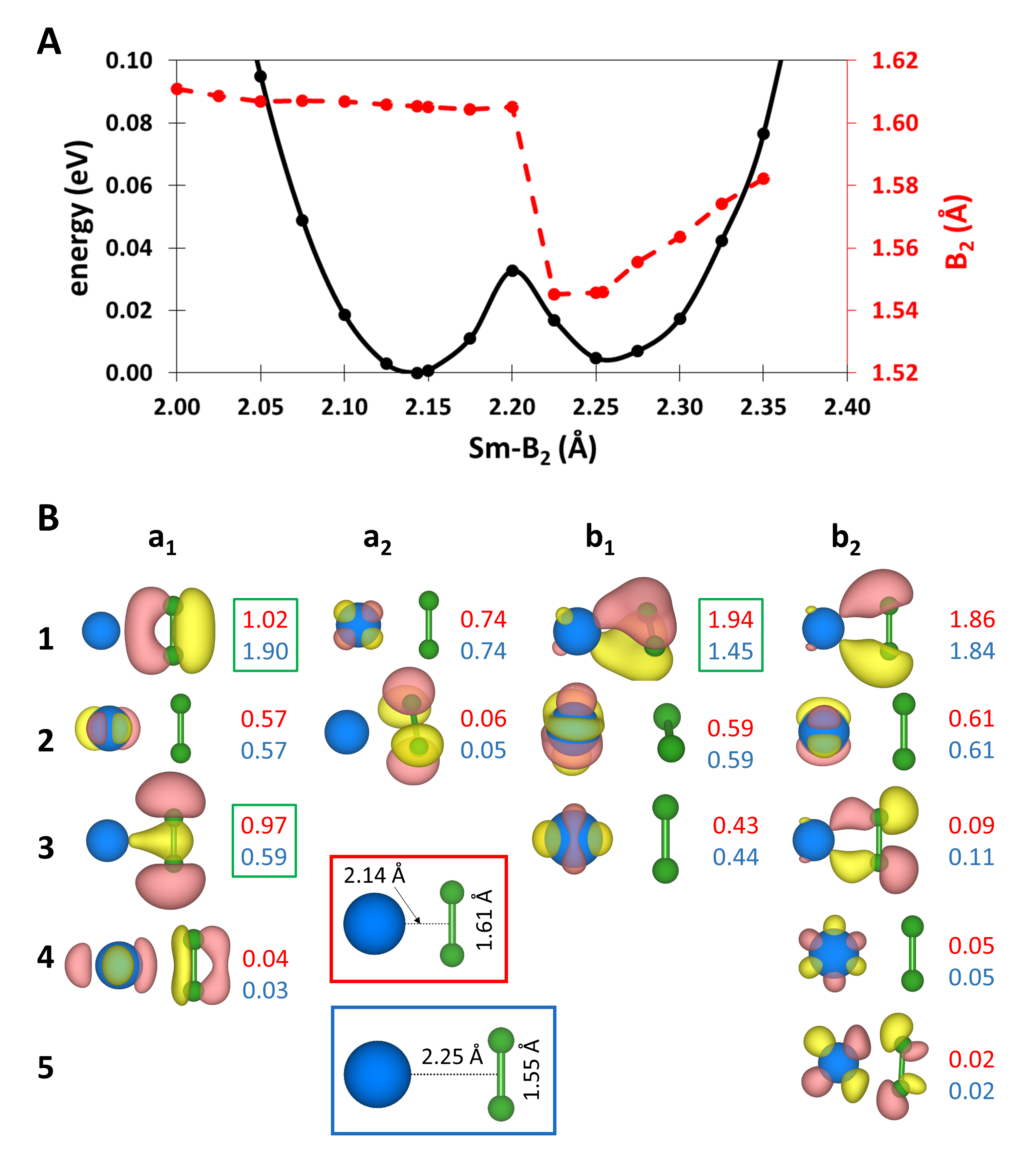}
\caption{
$\text{SmB}_2^+$ minimal cluster model. A: Ground state energy (black) and $\text{B}_2$ bond length (red). Calculations were carried out at the (8-SA-CASSCF(14,9)/MRCI + DKH6) level of theory, and provided an energy difference between both minima of 4 meV. This value is in agreement with that provided by SO(8-SA-CASSCF(14,9)/NEVPT2 + DKH6). Additional methodological details are provided in the Supporting Information. B: The active space orbitals, and the occupations for the two minima (geometries of the minima are shown in the insets: $\text{B}_2$ dimer at the center of the face in the cubic lattice---blue, Sm closer to $\text{B}_2$---red). The oxidation state of Sm switches between the two minima. Electrons from the Sm--$\text{B}_2$ bonding orbitals ($^3\text{a}_1$ and $^1\text{b}_1$) shift to the $\pi(p_z)$ bonding states of $\text{B}_2$ ($^1\text{a}_1$). These orbitals are highighted in green.
}
\label{fig1}
\end{figure}

Unexpectedly, the ground state potential energy surface of $\text{SmB}_2^+$ has two nearly degenerate bonding minima, one with the B-B bond length of 1.61 {\AA} and the Sm-$\text{B}_2$ distance of 2.14 {\AA}, and the other with the B-B bond length of 1.55 {\AA} and the Sm-$\text{B}_2$ distance of 2.25 {\AA} (Figure 1).  The energy difference between the minima is only 4 meV, suggestive of easy interconversion on short timescales. The two minima are predominantly $^6\text{A}_2$ and $^4\text{A}_2$ (see Figure 1A and Tables S1-S5 in Supporting Information). The electronic transition between them is achieved by nuclear motion, and enabled by strong nonadiabatic coupling. Most of the orbitals in the active space remain constantly occupied during the transition and in the minima (see Figure 1B), and many correspond to Sm-B bonds, directly contrasting the long-held presumption of boron's innocence. There is one crucial bonding difference between the minima: a significant shift of electrons from $^3\text{a}_1$ and $^1\text{b}_1$ to $^1\text{a}_1$ (highlighted in green in Figure 1), corresponding to one electron transfer from the covalent Sm–-$\text{B}_2$ orbitals to an isolated $\pi(p_z)$ bond on $\text{B}_2$. 
As a result, bonding orbitals $^3\text{a}_1$ and $^1\text{b}_1$ lose about half of an electron each, while the $\text{B}_2$ $^1\text{a}_1$ orbital gains one electron. According to our interpretation, the first minimum (short Sm--$\text{B}_2$ distance) has a covalent character, with Sm in the +2 oxidation state, and the second one (longer Sm--$\text{B}_2$) corresponds to a more ionic system, in which Sm is in the +3 oxidation state, as evidenced by the electron transference to the doubly occupied B-B $^1\text{a}_1$ orbital. This way, whichever of these orbitals is occupied will determine the oxidation state of Sm. Remarkably, the previous assignment of oxidation states is supported by the the analysis of the spin density. In particular, the Mulliken spin density over Sm in the first minimum (Sm$^{2+}$) is +3.47, while for the second minimum (Sm$^{3+}$) it is +2.42; pointing towards the first minimum losing one electron. This way, the small stretch and displacement of the $\text{B}_2$ causes the change of the electronic configuration and charge transfer. Notice that the difference in the Sm--$\text{B}_2$ distance between both minima is only 0.11 {\AA}, and the difference in B-B distances is 0.06 {\AA}. The oxidation state of Sm is thus controlled by motion of boron. In the Sm$\text{B}_6$ solid, this motion is achieved through vibrations.

Explicitly computing the two states and the vibronically-enabled transition between them in the solid is not feasible, 
but we gather significant evidence that the two minima are present. 
DFT provides a qualitative agreement: upon an artificial deformation of the lattice where $\text{B}_2$ is displaced toward one of the Sm ions by 20\% of the equilibrium Sm--$\text{B}_2$ distance, the $\text{B}_2$ $\pi(p_z)$ state moves from the valence to the virtual manifold, as predicted by our cluster model (Fig. S2). 
The average valence measured by M{\"o}ssbauer is static over changes in T, whereas X-ray absorption shows the increase of valence with T \cite{16mizumaki2009temperature}. Our model resolves this discrepancy by predicting a constant ratio of $f$-shell occupations (which M{\"o}ssbauer measures). Indeed, our minimal cluster model shows the $\text{Sm(II)} \to \text{Sm(III)}$ transition as an electron transference from Sm--$\text{B}_2$ bonding orbitals to $\text{B}_2$, enabled by boron motion. Hence, we have strong initial indications that the solid $\text{SmB}_6$ also has two distinct bonded states available to every $\text{B}_2$ with every Sm in the face of the cube. Based on this two-state model, we now construct the model of the full solid.

For the face of the cubic lattice, consisting of four Sm atoms and one $\text{B}_2$, we thus predict a 5-well system (Figure 2A), with four energy-degenerate wells where one out of four Sm ions is in the +2 oxidation state (reduced) and $\text{B}_2$ is stretched and displaced toward this Sm, and a non-degenerate central well (oxidized) where the B-B distance is slightly compressed, $\text{B}_2$ holds an extra electron in the bonding $\pi(p_z)$-orbital, and all Sm ions are in the +3 oxidation state (oxidized). The relative energies and geometries of these minima in the solid likely differ from those in the cluster. The energy-displacement between the minima presents three distinct possibilities for the character of the mixed valent system (Figure 2A right). First, the oxidized state can be lower in energy than the reduced states, and this case results in a symmetric oxidized ground state and thermal population of the reduced states. Second, the oxidized state can be higher in energy than the reduced states but lower than the barrier to interconversion between the states. This system would undergo transitions between the reduced minima with different Sm ions in the +2 state, and the ground state would be symmetry-broken. Finally, the oxidized minimum can be higher in energy than the interconversion barrier between the reduced wells. The mixed valency in this case will not be affected by the oxidized state, and it falls into one of the three Robin-Day schemes. 

The free parameter in the model is the energetic splitting $\epsilon$ between the reduced and oxidized minima. Instead of directly using $\epsilon$ from the cluster model, or calculating it for the solid ab inito, we find it from fitting to experiment, and then compare it with the value provided by the cluster model (4 meV). Given that only one Sm in every face of the cube can interact with $\text{B}_2$, we can state that, when $\text{B}_2$ interacts with the Sm, the average oxidation state of Sm in the solid will be +2.5, while when $\text{B}_2$ is non-interacting, the average oxidation state of Sm will be in the +3. The dimensionality of the problem is thus reduced to a set of cluster states embedded in the solid. Using x-ray absorption average valency data as a function of $\text{T,}^{16}$ we performed a least-squares fit to the functional form shown in Figure 2B, and found the energy splitting, $\epsilon$, in the solid to be 5.2 meV. The reduced minima appear lower in energy than the oxidized minimum. Notice that this value is in quantitative agreement with the energy difference between the ionic and covalent minima found in the cluster model. The fitted degeneracy of the reduced states, $d$, is 4.05 as expected for a four-degenerate system (Figure 2B).

The model yields the following description of mixed valence in $\text{SmB}_6$: Below 10 K, the solid is almost entirely in the ground state with an average valency of +2.5. The $\text{B}_2$ dimers hold no extra electrons, remain stretched, and displaced toward one of the Sm, creating structural disorder in the boron sublattice. Between ca. 10 K and 100 K, the thermal excitations begin allowing $\text{B}_2$ to exit the reduced minima, now holding an extra electron, and leaving Sm in the +3 oxidation state. Hence, valency increases rapidly until beginning to taper off towards an asymptotic average of +2.6. 

\begin{figure*}
\centering 
\includegraphics[width=0.65\linewidth]{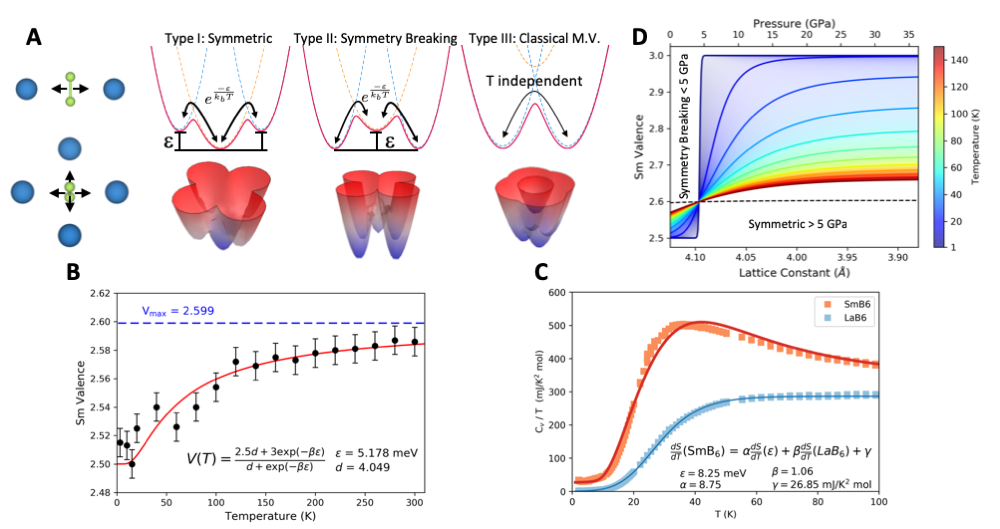}
\caption{Model of $\text{SmB}_6$ solid based on the dually accessible bonding states. A: top: one-dimensional three-well model displaying the three classes of mixed valency that can occur with a variable central well height and their thermodynamic populations. Bottom: two-dimensional analogues of the displacements and well types as they would be seen in the solid B: Least squares two-level thermodynamic fit to the x-ray absorption mixed valency data. The two free parameters d and $\epsilon$ are the degeneracy of the side states and the energy splitting respectively. C: Fitting of the excess entropy to the toy model. The modes from the rest of the solid are accounted for by scaling the profile of $\text{LaB}_6$ and adding the additional peak on top. D: Demonstration of a transition between mixed valency types induced by pressure calculated from x-ray data over different pressures.  Note the discontinuous jump of average Sm valence from +2.5 to +3 at 5 GPa signals a phase transition from symmetry broken to symmetric mixed valency. We used the parameters calculated in section B.}
\label{fig2}
\end{figure*}

Vibronic coupling thus appears to be the key to mixed valency in $\text{SmB}_6$. Evidence of the electron-phonon interactions between the B-B stretch in the $\text{B}_2$ dimers and the electrons in Sm comes from Raman vibrational spectroscopy.

Guided by the model, we tested the dependence of Raman spectra of $\text{SmB}_6$ on the ${^{10}\text{B}} / {^{11}\text{B}}$ isotope substitution. Notice that the experimental details details are provided in the SI. The spectra show a broad continuum of excitations extending to frequencies above 200 meV, with narrow features of phonons superimposed on it at frequencies between approximately 80 and 175 meV (Figure 3A). The continuum is assigned to the excitations between the bands of Sm, which develop a hybridization gap at low temperatures \cite{34valentine2016breakdown}.  The phonons are Raman active boron-motion related phonons, which shift in frequency upon isotope substitution from $^{10}\text{B}$ to $^{11}\text{B}$ as expected according to the regular dependence of frequency on mass as 
$\omega \sim \sqrt{\dfrac{m(^{11}\text{B})}{m(^{10}\text{B})}}$.
The line width of the phonons that do not interact with electrons, $\Gamma$, is defined by a natural width $\Gamma_N$ for pure isotopes, and $\Gamma_N + \Gamma_D$ for partial isotope substitution. $\Gamma_D$ depends on disorder (Figure 3B), as demonstrated by the $E_g$ and $T_{1g}$ phonons, which show an increased width for samples with mixed isotope content. The highest frequency $\text{A}_{1\text{g}}$ phonon is an exception from this rule. It shows an asymmetric so-called Fano shape [see SI], typical for phonons coupled to an underlying continuum of interband excitations of electrons of Sm\cite{35fano1961effects}.  The constant width $\Gamma$ of $\text{A}_{1\text{g}}$ for all levels of isotope substitution suggests that it is defined by the electron-phonon coupling. This strong electron-phonon interaction is in agreement with our model, where an in-phase stretch of the B-B bonds in the $\text{B}_2$ dimers (Figure 3D) is associated with Sm-$\text{B}_2$ electronic interactions responsible for the lowering of the oxidation state of the Sm. 

\begin{figure*}[t]
\centering 
\includegraphics[width=0.65\linewidth]{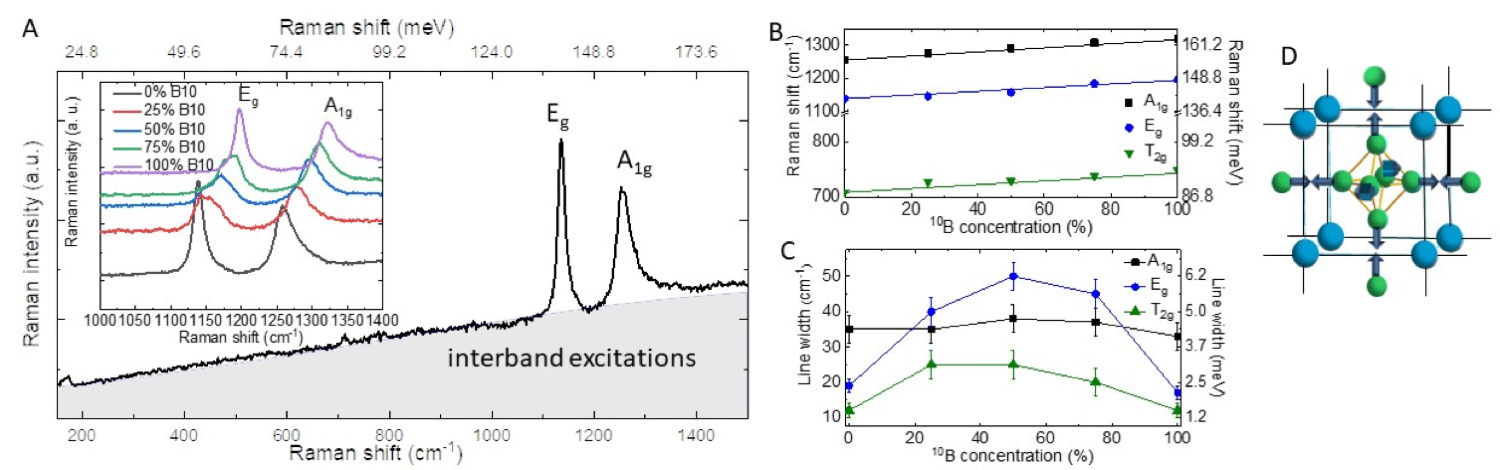}
\caption{Electron-phonon coupling revealed by boron-motion-related phonons of $\text{SmB}_6$  A: Raman scattering spectrum of $\text{Sm}^{11}\text{B}_6$ in $\text{A}_{1\text{g}} + \text{E}_{\text{g}}$ symmetry. Phonons are superimposed on the continuum of interband excitations. $\text{A}_{1\text{g}}$ phonon shows an asymmetric shape due to the interactions with the interband excitations. Inset shows phonon spectra of isotope substituted $\text{SmB}_6$ for $\text{A}_{1\text{g}} + \text{E}_{\text{g}}$ symmetry, shifted along Y axis for clarity; note the linear increase of frequency upon changing from $^{11}$B to $^{10}$B (plotted in B), and an increase of the width of the Eg and T2g phonons for partially substituted samples (plotted in C). D: The $\text{A}_{1\text{g}}$ phonon of $\text{SmB}_6$ following calculations presented in Ref. (\citenum{36ogita2003raman} ). Note that it involves in-phase stretching vibrations of all B-B pairs.}
\label{fig3}
\end{figure*}

Raman spectra of $\text{SmB}_6$ also show a number of features in the energy range of interest identified by our model (Figure 4A): We observe two overlapping features  at ca. 22 meV marked as F1 in Figure 4, one of which is a sharp phonon-like (red circle), and the other is much broader (black square). Lattice optical $\text{T}_{1\text{u}}$ phonons involving movement of both boron and Sm are good candidates for an excitation observed in this frequency range\cite{36ogita2003raman,37alekseev2015high}. This assignment is confirmed by the boron isotope-dependent frequency shift of the sharp feature.  Indeed, while low frequency of around 20 meV where the phonon is observed is defined by the large mass of Sm atom, the feature shifts by ca. 0.17 meV towards the higher frequency upon replacement of $^{11}\text{B}$ with $^{10}\text{B}$ (Figure 4A). This phonon modulates the distance between $\text{B}_2$ and Sm atoms, which within our model leads to valence fluctuations. In the absence of electron-phonon coupling $\text{T}_{1\text{u}}$ phonons are forbidden in Raman scattering of $\text{SmB}_6$. Valence fluctuations lead to relaxing of the selection rules, and to an appearance of an additional broad excitation which is associated with a local lattice deformation and a transfer of an electron from Sm to $\text{B}_2$. 

While our model provides for the first time a detailed microscopic description of the valence fluctuations phenomena in $\text{SmB}_6$, the general picture resonates with the theory of exiton-polaron excitations developed in Ref. \citenum{37alekseev2015high}-\citenum{38erten2016kondo}. 
The $\text{A}_{1\text{g}}$ symmetry of the features at about 20 meV (Figure 4D) is in agreement with the suggested symmetry of the valence fluctuations\cite{37alekseev2015high,38erten2016kondo}. Below 50 K the linewidth of the narrow feature of the $\text{T}_{1\text{u}}$ phonon starts to decrease, while the feature shifts to higher frequencies (Figure 4C), evidencing that the system descends more deeply in the reduced minima, away from the unharmonic regime and strongest vibronic coupling. The behavior follows the suggestion of our model, where the system freezes at the bottom of the reduced minima at low T, leading to the +2.5 average valency of Sm. On the other hand, the mixed exiton-polaron excitation shows lower bandwidth in the higher-temperature regime, when the exciton-polaron is in a hopping regime and is suggested to have longer life-time in each site\cite{38erten2016kondo}. 

\begin{figure*}[t]
\centering 
\includegraphics[width=0.65\linewidth]{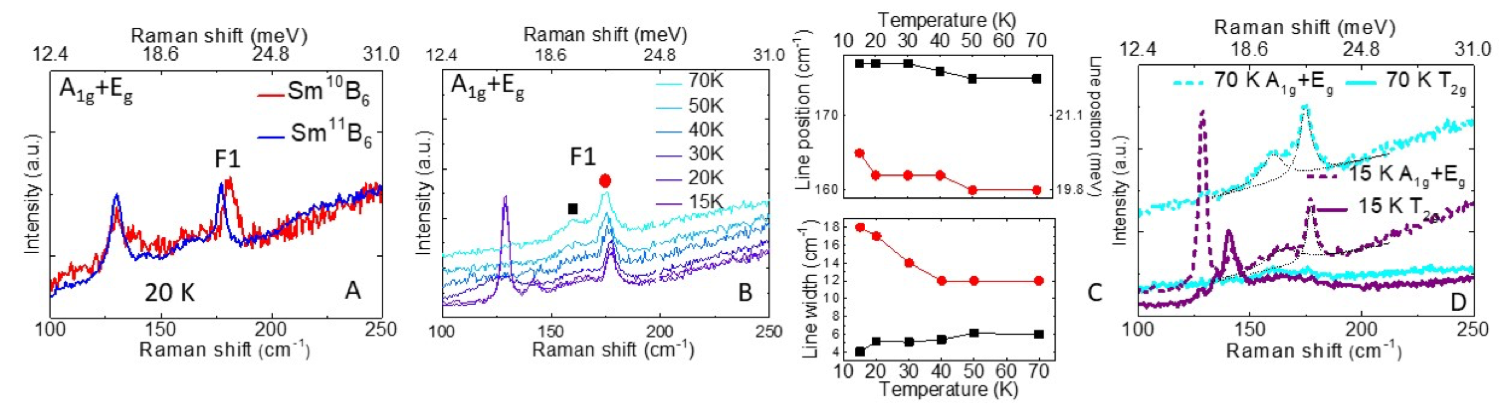}
\caption{Raman spectra in the range between 12.4 and 30 meV where valence fluctuations-related feature (F1) is observed. A: Spectra of $\text{SmB}_6$ with fully pure $^{10}$B and $^{11}$B isotopes in $\text{A}_{1\text{g}}+\text{E}_\text{g}$ scattering channel at 20 K. Note that isotope shift is observed for the feature related to optical phonon at about 22 meV, while position of spin exciton at 16 meV does not change. B-C: $\text{SmB}_6$ (natural isotope mix) $\text{A}_{1\text{g}}+\text{E}_\text{g}$ scattering channel, temperature dependence of the spectra and lines positions and width for features related to valence fluctuations. Note that the broad exiton-polaron excitation broadens considerably on cooing below 40 K, while the narrow phonon-related feature gets somewhat narrower. D: Comparison of $\text{A}_{1\text{g}}+\text{E}_\text{g}$ and $\text{T}_{1\text{g}}$ symmetry spectra. Fitting curves for the lattice phonon and exciton-polaron excitations are shown for 15 and 70 K with black dashed curves. Note that the features related to valence fluctuations are observed only in $\text{A}_{1\text{g}}$.}
\label{fig4}
\end{figure*}

Hence, both theory and spectroscopy point at rich vibronic behavior of $\text{SmB}_6$. On the basis of this understanding, we now address the question of why $\text{SmB}_6$ has an anomalous peak in the specif heat at constant volume ($\text{C}_V$) at around 20 K \cite{15phelan2014correlation}. Normally, one would expect $\left. \dfrac{dS}{dT} \right|_V$, which is equal to $\text{C}_V$/T, to decrease with T, as the phonons gradually freeze out. The additional peak points towards an unexplained energy scale in $\text{SmB}_6$. From our model, we derive the expression for the entropy and calculate its maximum at 19.6 K---in agreement with experiment (derivations presented in the SI). We interpret the anomalous peak in entropy as follows: at low T, vibrational entropy builds up from the motion of the $\text{B}_2$ units within the reduced minima; as T rises, $\text{B}_2$ units unfreeze from the reduced minima and gain a new vibrational freedom, the lattice changes symmetry, and the vibrational partition function qualitatively changes its nature. Here we obtain a second, and independent, method of finding the energy splitting, $\epsilon$, (Figure 2C, more details are provided in the SI). Specifically, we find quantitative agreement with the experiment for the excess entropy when $\epsilon = 8.25$ meV, similar to the previous estimations, with the difference likely arising from the simplicity of the model.

The average valency in $\text{SmB}_6$ also responds to the pressure. With increasing p, it increases past what would be allowed by thermal population, e.g. at 35 GPa the average valency approaches +3\cite{17butch2016pressure}. Curiously, the average valency never reaches +3 nor is it affected by a phase transition from a non-magnetic to magnetic material at 5 GPa. Indeed, as Butch et all point out \cite{17butch2016pressure}, this result challenges prior explanations of intermediate valence in $\text{SmB}_6$ that do not predict the mixed valency remaining stable at high pressure. Microscopically, pressure shrinks the unit cell and thus crowds the Sm closer to the $\text{B}_2$ units. From the cluster model, we see that the oxidized Sm-$\text{B}_2$ minimum is more easily compressed than the reduced minimum, with the force constants of 9.5 eV\AA$^{-2}$ and 10.4 eV\AA$^{-2}$, respectively. Hence, at rising p, the reduced minima rise in energy, eventually going above the oxidized minimum, which becomes the new ground state, leading to a phase transition. 

In order to test this hypothesis, we calculate the energy splitting $\epsilon$ for each experimentally tested p, and found that it decreases with added p. At 5 GPa the material undergoes a first order phase transition from the system where the central oxidized state mediates the transitions between the reduced states through boron motion, to the system where the oxidized state is lower in energy and dominates, as evidenced by a discontinuity in the average valency (Figure 2D). This lends an explanation to the observed transition from non-magnetic to magnetic solid at that pressure: the conducting and magnetic phase corresponds to the oxidized state, where boron holds an extra electron and thus a spin. This also provides an answer for why below 5 GPa, the solid has a ground state average valency of +2.5 while above 5 GPa the solid has a ground state average valency near +3. However, for any $\text{T} > 0\text{ K}$, adding p will never cause the average valency to reach +3, due to thermal fluctuations and transient visits of boron closer to Sm. It's worth noting that at $\text{k}_\text{b}\text{T} \gg \epsilon$ the phase transition is indiscernible.

The two different $\text{Sm-B}_2$ bonding modes lead to the re-interpretation of the photoelectron spectrum of $\text{SmB}_6$ \cite{Chazalviel}. There are two sets of photodetachment peaks, at 0-1.5 eV and 5-12 eV, traditionally assigned based on the atomic spectra of $\text{Sm}^{2+}$ and $\text{Sm}^{3+}$, respectively. However, Sm engages in covalent bonding with the B, and cannot be viewed as an isolated ion. Since in the oxidized minimum, Sm is +3, and $\text{B}_2$ holds an electron, we computed the vertical detachment energies (VDE) of $\text{B}_2^{-}$ (doublet) to be 1.02 eV for the triplet final state, and 1.89 eV for the singlet. These values fit the low-energy part of the spectrum classically assigned to $\text{Sm}^{2+}$ transitions. The deeper peaks originate from the $\text{Sm-B}_2$ cluster, which is the quasi-isolated unit in $\text{SmB}_6$ instead of $\text{Sm}^{2+/3+}$. The two minima are close in energy, and both yield peaks in the same energy range (Tables SVI, SVII). Hence, we propose that the $\text{Sm-B}_2$ cluster with Sm in both oxidation states contribute to the deeper peaks in the spectra.

Finally, we consider the outstanding problem of the quantum oscillations in $\text{SmB}_6$. 
Both 2D and 3D Fermi surfaces were reported for this insulating solid, although these conclusions are model-dependent and based on very similar data
\cite{38erten2016kondo,39hartstein2018fermi}. 
Currently, the origin of quantum oscillations is understood as composite particles with zero overall charge and a non-zero spin. 
From our model, we propose a potential physical origin for this many-body effect:
every Sm ion is in an electron-exchange relationship with one $\mathrm{B}_2$
(we showed in a cluster model that every Sm can sustain the tighter bonded state with only one $\text{B}_2$ at a time). The full stoichiometric unit is required, however, for the charge neutrality. 
The two possible minima each have a different overall spin and each of these spin 
states has $2s+1$ degenerate magnetic states. 
At finite magnetic fields, the Zeeman splitting  breaks the degeneracy of the magnetic states and state-state couplings lead to avoided crossings. 
These non-linearities in state energy as a function of magnetic field 
introduce deformities in the Helmholtz free energy and necessarily lead to
a non-monotonic magnetic susceptibility, explaining the ambiguous magnetic Fermi surface
(full discussion presented in SI).

\section*{Conclusion}

In summary, we present a new paradigm for $\text{SmB}_6$ in which strong vibronic effects allow the coexistence of five distinct bonding possibilities for each $\text{B}_2$ unit connecting the neighboring unit cells. Through this new model we discovered a new class of mixed valency, distinct from the Robin-Day scheme; rather than being dependent on couplings of two reduced states, the mixed valency results from the population of a third, central, oxidized state. The existence of the extra states in the material, and the vibronic coupling between them, enabled by the boron motion, creates an additional, lower energy scale to the solid, accounting for excess entropy and an unexplained phase transition. Hand in hand, this model also suggests that the solid has hidden aperiodicity within it; namely, the $\text{B}_2$ units are disordered within the cubic Sm lattice at low T. This realization naturally leads to an explanation of why symmetry forbidden peaks persistently appear in Raman spectra of $\text{SmB}_6$. We also propose these cluster states to be the physical manifestations of theorized composite particles responsible for the observed magnetic Fermi surface.

\section*{Acknowledgements}
A.N.A. acknowledges the support of the NSF CAREER Award (CHE-1351968). P.J.R. acknowledges support from the NSF Graduate Research Fellowship under Grant No. (DGE-1644869), and the UCLA Undergraduate Research Center Sciences. Work at the Institute for Quantum Matter, an Energy Frontier Research Center, was funded by the U.S. Department of Energy, Office of Science, Office of Basic Energy Sciences, under Award DE-SC0019331. P. F. S. R. acknowledges support from the Laboratory Directed Research and Development program under project No. 20180618ECR.

\bibliography{citations}

\begin{thebibliography}{10}
\expandafter\ifx\csname url\endcsname\relax
  \def\url#1{\texttt{#1}}\fi
\expandafter\ifx\csname urlprefix\endcsname\relax\def\urlprefix{URL }\fi
\providecommand{\bibinfo}[2]{#2}
\providecommand{\eprint}[2][]{\url{#2}}

\bibitem{1nickerson1971physical}
\bibinfo{author}{Nickerson, J.} \emph{et~al.}
\newblock \bibinfo{title}{Physical properties of $\text{SmB}_6$}.
\newblock \emph{\bibinfo{journal}{Phys. Rev. B}} \textbf{\bibinfo{volume}{3}},
  \bibinfo{pages}{2030} (\bibinfo{year}{1971}).

\bibitem{2cooley1995sm}
\bibinfo{author}{Cooley, J.}, \bibinfo{author}{Aronson, M.},
  \bibinfo{author}{Fisk, Z.} \& \bibinfo{author}{Canfield, P.}
\newblock \bibinfo{title}{$\text{SmB}_6$: Kondo insulator or exotic metal?}
\newblock \emph{\bibinfo{journal}{Phys. Rev. Lett.}}
  \textbf{\bibinfo{volume}{74}}, \bibinfo{pages}{1629} (\bibinfo{year}{1995}).

\bibitem{3martin1979theory}
\bibinfo{author}{Martin, R.~M.} \& \bibinfo{author}{Allen, J.}
\newblock \bibinfo{title}{Theory of mixed valence: Metals or small gap
  insulators}.
\newblock \emph{\bibinfo{journal}{J. Appl. Phys.}}
  \textbf{\bibinfo{volume}{50}}, \bibinfo{pages}{7561--7566}
  (\bibinfo{year}{1979}).

\bibitem{4li2014two}
\bibinfo{author}{Li, G.} \emph{et~al.}
\newblock \bibinfo{title}{Two-dimensional fermi surfaces in kondo insulator
  $\text{SmB}_6$}.
\newblock \emph{\bibinfo{journal}{Science}} \textbf{\bibinfo{volume}{346}},
  \bibinfo{pages}{1208--1212} (\bibinfo{year}{2014}).

\bibitem{5jiang2013observation}
\bibinfo{author}{Jiang, J.} \emph{et~al.}
\newblock \bibinfo{title}{Observation of possible topological in-gap surface
  states in the kondo insulator $\text{SmB}_6$ by photoemission}.
\newblock \emph{\bibinfo{journal}{Nat. Commun.}} \textbf{\bibinfo{volume}{4}},
  \bibinfo{pages}{3010} (\bibinfo{year}{2013}).

\bibitem{6stern2017surface}
\bibinfo{author}{Stern, A.}, \bibinfo{author}{Dzero, M.},
  \bibinfo{author}{Galitski, V.}, \bibinfo{author}{Fisk, Z.} \&
  \bibinfo{author}{Xia, J.}
\newblock \bibinfo{title}{Surface-dominated conduction up to 240 k in the kondo
  insulator $\text{SmB}_6$ under strain}.
\newblock \emph{\bibinfo{journal}{Nat. Mater.}} \textbf{\bibinfo{volume}{16}},
  \bibinfo{pages}{708} (\bibinfo{year}{2017}).

\bibitem{7kim2014topological}
\bibinfo{author}{Kim, D.-J.}, \bibinfo{author}{Xia, J.} \&
  \bibinfo{author}{Fisk, Z.}
\newblock \bibinfo{title}{Topological surface state in the kondo insulator
  samarium hexaboride}.
\newblock \emph{\bibinfo{journal}{Nat. Mater.}} \textbf{\bibinfo{volume}{13}},
  \bibinfo{pages}{466} (\bibinfo{year}{2014}).

\bibitem{8park2016topological}
\bibinfo{author}{Park, W.~K.} \emph{et~al.}
\newblock \bibinfo{title}{Topological surface states interacting with bulk
  excitations in the kondo insulator $\text{SmB}_6$ revealed via planar
  tunneling spectroscopy}.
\newblock \emph{\bibinfo{journal}{Proc. Natl. Acad. Sci. U.S.A.}}
  \textbf{\bibinfo{volume}{113}}, \bibinfo{pages}{6599--6604}
  (\bibinfo{year}{2016}).

\bibitem{9tan2015unconventional}
\bibinfo{author}{Tan, B.} \emph{et~al.}
\newblock \bibinfo{title}{Unconventional fermi surface in an insulating state}.
\newblock \emph{\bibinfo{journal}{Science}} \textbf{\bibinfo{volume}{349}},
  \bibinfo{pages}{287--290} (\bibinfo{year}{2015}).

\bibitem{10hartstein2018fermi}
\bibinfo{author}{Hartstein, M.} \emph{et~al.}
\newblock \bibinfo{title}{Fermi surface in the absence of a fermi liquid in the
  kondo insulator $\text{SmB}_6$}.
\newblock \emph{\bibinfo{journal}{Nat. Phys.}} \textbf{\bibinfo{volume}{14}},
  \bibinfo{pages}{166} (\bibinfo{year}{2018}).

\bibitem{11knolle2017excitons}
\bibinfo{author}{Knolle, J.} \& \bibinfo{author}{Cooper, N.~R.}
\newblock \bibinfo{title}{Excitons in topological kondo insulators: Theory of
  thermodynamic and transport anomalies in $\text{SmB}_6$}.
\newblock \emph{\bibinfo{journal}{Phys. Rev. Lett.}}
  \textbf{\bibinfo{volume}{118}}, \bibinfo{pages}{096604}
  (\bibinfo{year}{2017}).

\bibitem{12baskaran2015majorana}
\bibinfo{author}{Baskaran, G.}
\newblock \bibinfo{title}{Majorana fermi sea in insulating $\text{SmB}_6$: A
  proposal and a theory of quantum oscillations in kondo insulators}
  (\bibinfo{year}{2015}).
\newblock \bibinfo{note}{Preprint at https://arxiv.org/abs/1507.03477}.

\bibitem{13erten2017skyrme}
\bibinfo{author}{Erten, O.}, \bibinfo{author}{Chang, P.-Y.},
  \bibinfo{author}{Coleman, P.} \& \bibinfo{author}{Tsvelik, A.~M.}
\newblock \bibinfo{title}{Skyrme insulators: insulators at the brink of
  superconductivity}.
\newblock \emph{\bibinfo{journal}{Phys. Rev. Lett.}}
  \textbf{\bibinfo{volume}{119}}, \bibinfo{pages}{057603}
  (\bibinfo{year}{2017}).

\bibitem{14dzero2016topological}
\bibinfo{author}{Dzero, M.}, \bibinfo{author}{Xia, J.},
  \bibinfo{author}{Galitski, V.} \& \bibinfo{author}{Coleman, P.}
\newblock \bibinfo{title}{Topological kondo insulators}.
\newblock \emph{\bibinfo{journal}{Annu. Rev. Condens. Matter Phys.}}
  \textbf{\bibinfo{volume}{7}}, \bibinfo{pages}{249--280}
  (\bibinfo{year}{2016}).

\bibitem{15phelan2014correlation}
\bibinfo{author}{Phelan, W.} \emph{et~al.}
\newblock \bibinfo{title}{Correlation between bulk thermodynamic measurements
  and the low-temperature-resistance plateau in $\text{SmB}_6$}.
\newblock \emph{\bibinfo{journal}{Phys. Rev. X}} \textbf{\bibinfo{volume}{4}},
  \bibinfo{pages}{031012} (\bibinfo{year}{2014}).

\bibitem{16mizumaki2009temperature}
\bibinfo{author}{Mizumaki, M.}, \bibinfo{author}{Tsutsui, S.} \&
  \bibinfo{author}{Iga, F.}
\newblock \bibinfo{title}{Temperature dependence of sm valence in
  $\text{SmB}_6$ studied by x-ray absorption spectroscopy}.
\newblock \emph{\bibinfo{journal}{J. Phys. Conf. Ser.}}
  \textbf{\bibinfo{volume}{176}}, \bibinfo{pages}{012034}
  (\bibinfo{year}{2009}).

\bibitem{17butch2016pressure}
\bibinfo{author}{Butch, N.~P.} \emph{et~al.}
\newblock \bibinfo{title}{Pressure-resistant intermediate valence in the kondo
  insulator $\text{SmB}_6$}.
\newblock \emph{\bibinfo{journal}{Phys. Rev. Lett.}}
  \textbf{\bibinfo{volume}{116}}, \bibinfo{pages}{156401}
  (\bibinfo{year}{2016}).

\bibitem{18parthey2014quantum}
\bibinfo{author}{Parthey, M.} \& \bibinfo{author}{Kaupp, M.}
\newblock \bibinfo{title}{Quantum-chemical insights into mixed-valence systems:
  within and beyond the robin--day scheme}.
\newblock \emph{\bibinfo{journal}{Chem. Soc. Rev.}}
  \textbf{\bibinfo{volume}{43}}, \bibinfo{pages}{5067--5088}
  (\bibinfo{year}{2014}).

\bibitem{19robinson1979valence}
\bibinfo{author}{Robinson, J.~M.}
\newblock \bibinfo{title}{Valence transitions and intermediate valence states
  in rare earth and actinide materials}.
\newblock \emph{\bibinfo{journal}{Phys. Rep.}} \textbf{\bibinfo{volume}{51}},
  \bibinfo{pages}{1--62} (\bibinfo{year}{1979}).

\bibitem{20ramirez1970metal}
\bibinfo{author}{Ramirez, R.}, \bibinfo{author}{Falicov, L.} \&
  \bibinfo{author}{Kimball, J.}
\newblock \bibinfo{title}{Metal-insulator transitions: A simple theoretical
  model}.
\newblock \emph{\bibinfo{journal}{Phys. Rev. B}} \textbf{\bibinfo{volume}{2}},
  \bibinfo{pages}{3383} (\bibinfo{year}{1970}).

\bibitem{21levine2009advancements}
\bibinfo{author}{Levine, J.~B.}, \bibinfo{author}{Tolbert, S.~H.} \&
  \bibinfo{author}{Kaner, R.~B.}
\newblock \bibinfo{title}{Advancements in the search for superhard
  ultra-incompressible metal borides}.
\newblock \emph{\bibinfo{journal}{Adv. Funct. Mater.}}
  \textbf{\bibinfo{volume}{19}}, \bibinfo{pages}{3519--3533}
  (\bibinfo{year}{2009}).

\bibitem{22gilman2006design}
\bibinfo{author}{Gilman, J.}, \bibinfo{author}{Cumberland, R.} \&
  \bibinfo{author}{Kaner, R.}
\newblock \bibinfo{title}{Design of hard crystals}.
\newblock \emph{\bibinfo{journal}{Int. J. Refract. Metals. Hard. Mater.}}
  \textbf{\bibinfo{volume}{24}}, \bibinfo{pages}{1--5} (\bibinfo{year}{2006}).

\bibitem{23brazhkin2002harder}
\bibinfo{author}{Brazhkin, V.~V.}, \bibinfo{author}{Lyapin, A.~G.} \&
  \bibinfo{author}{Hemley, R.~J.}
\newblock \bibinfo{title}{Harder than diamond: dreams and reality}.
\newblock \emph{\bibinfo{journal}{Philos. Mag. A}}
  \textbf{\bibinfo{volume}{82}}, \bibinfo{pages}{231--253}
  (\bibinfo{year}{2002}).

\bibitem{24robinson2017mystery}
\bibinfo{author}{Robinson, P.~J.} \emph{et~al.}
\newblock \bibinfo{title}{Mystery of three borides: Differential metal--boron
  bonding governing superhard structures}.
\newblock \emph{\bibinfo{journal}{Chem. Mater.}} \textbf{\bibinfo{volume}{29}},
  \bibinfo{pages}{9892--9896} (\bibinfo{year}{2017}).

\bibitem{25lei2018understanding}
\bibinfo{author}{Lei, J.} \emph{et~al.}
\newblock \bibinfo{title}{Understanding how bonding controls strength
  anisotropy in hard materials by comparing the high-pressure behavior of
  orthorhombic and tetragonal tungsten monoboride}.
\newblock \emph{\bibinfo{journal}{J. Phys. Chem. C}}
  \textbf{\bibinfo{volume}{122}}, \bibinfo{pages}{5647--5656}
  (\bibinfo{year}{2018}).

\bibitem{26pauling1948metallic}
\bibinfo{author}{Pauling, L.}
\newblock \bibinfo{title}{The metallic state}.
\newblock \emph{\bibinfo{journal}{Nature}} \textbf{\bibinfo{volume}{161}},
  \bibinfo{pages}{1019} (\bibinfo{year}{1948}).

\bibitem{27pauling1949resonating}
\bibinfo{author}{Pauling, L.~C.}
\newblock \bibinfo{title}{A resonating-valence-bond theory of metals and
  intermetallic compounds}.
\newblock \emph{\bibinfo{journal}{Proc. R. Soc. Lond. A. Math Phys. Sci.}}
  \textbf{\bibinfo{volume}{196}}, \bibinfo{pages}{343--362}
  (\bibinfo{year}{1949}).

\bibitem{28hoffmann1987chemistry}
\bibinfo{author}{Hoffmann, R.}
\newblock \bibinfo{title}{How chemistry and physics meet in the solid state}.
\newblock \emph{\bibinfo{journal}{Angew. Chem. Int. Ed. Engl.}}
  \textbf{\bibinfo{volume}{26}}, \bibinfo{pages}{846--878}
  (\bibinfo{year}{1987}).

\bibitem{29BurdettChemicalBonding}
\bibinfo{author}{Burdett, J.~K.}
\newblock \emph{\bibinfo{title}{Chemical Bonding in Solids}}
  (\bibinfo{publisher}{Wiley}, \bibinfo{year}{1997}).

\bibitem{30cox1987electronic}
\bibinfo{author}{Cox, P.~A.}
\newblock \emph{\bibinfo{title}{The electronic structure and chemistry of
  solids}} (\bibinfo{publisher}{Oxford univ. press Oxford etc.},
  \bibinfo{year}{1987}).

\bibitem{31canadellOrbtial}
\bibinfo{author}{Canadell, E.}, \bibinfo{author}{Doublet, M.-L.} \&
  \bibinfo{author}{Christophe, I.}
\newblock \emph{\bibinfo{title}{Orbital Approach to the Electronic Structure of
  Solids}} (\bibinfo{publisher}{Oxford univ. Press}, \bibinfo{year}{2012}).

\bibitem{32jiao2016additional}
\bibinfo{author}{Jiao, L.} \emph{et~al.}
\newblock \bibinfo{title}{Additional energy scale in $\text{SmB}_6$ at
  low-temperature}.
\newblock \emph{\bibinfo{journal}{Nat. Commun.}} \textbf{\bibinfo{volume}{7}},
  \bibinfo{pages}{13762} (\bibinfo{year}{2016}).

\bibitem{33robinson2017smb6}
\bibinfo{author}{Robinson, P.~J.}, \bibinfo{author}{Zhang, X.},
  \bibinfo{author}{McQueen, T.}, \bibinfo{author}{Bowen, K.~H.} \&
  \bibinfo{author}{Alexandrova, A.~N.}
\newblock \bibinfo{title}{$\text{SmB}_6^-$ cluster anion: Covalency involving f
  orbitals}.
\newblock \emph{\bibinfo{journal}{J. Phys. Chem. A}}
  \textbf{\bibinfo{volume}{121}}, \bibinfo{pages}{1849--1854}
  (\bibinfo{year}{2017}).

\bibitem{34valentine2016breakdown}
\bibinfo{author}{Valentine, M.~E.} \emph{et~al.}
\newblock \bibinfo{title}{Breakdown of the kondo insulating state in
  $\text{SmB}_6$ by introducing sm vacancies}.
\newblock \emph{\bibinfo{journal}{Phys. Rev. B}} \textbf{\bibinfo{volume}{94}},
  \bibinfo{pages}{075102} (\bibinfo{year}{2016}).

\bibitem{35fano1961effects}
\bibinfo{author}{Fano, U.}
\newblock \bibinfo{title}{Effects of configuration interaction on intensities
  and phase shifts}.
\newblock \emph{\bibinfo{journal}{Phys. Rev.}} \textbf{\bibinfo{volume}{124}},
  \bibinfo{pages}{1866} (\bibinfo{year}{1961}).

\bibitem{36ogita2003raman}
\bibinfo{author}{Ogita, N.} \emph{et~al.}
\newblock \bibinfo{title}{Raman scattering investigation of $\text{RB}_6$ ({R}=
  {Ca}, {La}, {Ce}, {Pr}, {Sm}, {Gd}, {Dy}, and {Yb})}.
\newblock \emph{\bibinfo{journal}{Phys. Rev. B}} \textbf{\bibinfo{volume}{68}},
  \bibinfo{pages}{224305} (\bibinfo{year}{2003}).

\bibitem{37alekseev2015high}
\bibinfo{author}{Alekseev, P.~A.}
\newblock \bibinfo{title}{High borides: determining the features and details of
  lattice dynamics from neutron spectroscopy}.
\newblock \emph{\bibinfo{journal}{Phys.-Uspekhi}}
  \textbf{\bibinfo{volume}{58}}, \bibinfo{pages}{330} (\bibinfo{year}{2015}).

\bibitem{38erten2016kondo}
\bibinfo{author}{Erten, O.}, \bibinfo{author}{Ghaemi, P.} \&
  \bibinfo{author}{Coleman, P.}
\newblock \bibinfo{title}{Kondo breakdown and quantum oscillations in
  $\text{SmB}_6$}.
\newblock \emph{\bibinfo{journal}{Phys. Rev. Lett.}}
  \textbf{\bibinfo{volume}{116}}, \bibinfo{pages}{046403}
  (\bibinfo{year}{2016}).

\bibitem{Chazalviel}
\bibinfo{author}{Chazalviel, J.~N.}, \bibinfo{author}{Campagna, M.},
  \bibinfo{author}{Wertheim, G.~K.} \& \bibinfo{author}{Schmidt, P.~H.}
\newblock \bibinfo{title}{Study of valence mixing in $\text{SmB}_6$ by x-ray
  photoelectron spectroscopy}.
\newblock \emph{\bibinfo{journal}{Phys. Rev. B}} \textbf{\bibinfo{volume}{14}},
  \bibinfo{pages}{4586--4592} (\bibinfo{year}{1976}).

\bibitem{39hartstein2018fermi}
\bibinfo{author}{Hartstein, M.} \emph{et~al.}
\newblock \bibinfo{title}{Fermi surface in the absence of a fermi liquid in the
  kondo insulator $\text{SmB}_6$}.
\newblock \emph{\bibinfo{journal}{Nat. Phys.}} \textbf{\bibinfo{volume}{14}},
  \bibinfo{pages}{166} (\bibinfo{year}{2018}).

\end{thebibliography}


\begin{thebibliography}{10}
\expandafter\ifx\csname url\endcsname\relax
  \def\url#1{\texttt{#1}}\fi
\expandafter\ifx\csname urlprefix\endcsname\relax\def\urlprefix{URL }\fi
\providecommand{\bibinfo}[2]{#2}
\providecommand{\eprint}[2][]{\url{#2}}

\bibitem{s1werner2012molpro}
\bibinfo{author}{Werner, H.-J.}, \bibinfo{author}{Knowles, P.~J.},
  \bibinfo{author}{Knizia, G.}, \bibinfo{author}{Manby, F.~R.} \&
  \bibinfo{author}{Sch{\"u}tz, M.}
\newblock \bibinfo{title}{Molpro: a general-purpose quantum chemistry program
  package}.
\newblock \emph{\bibinfo{journal}{Wiley Interdiscip. Rev. Comput. Mol. Sci}}
  \textbf{\bibinfo{volume}{2}}, \bibinfo{pages}{242--253}
  (\bibinfo{year}{2012}).

\bibitem{s2lu2016correlation}
\bibinfo{author}{Lu, Q.} \& \bibinfo{author}{Peterson, K.~A.}
\newblock \bibinfo{title}{Correlation consistent basis sets for lanthanides:
  The atoms la--lu}.
\newblock \emph{\bibinfo{journal}{J. Chem. Phys.}}
  \textbf{\bibinfo{volume}{145}}, \bibinfo{pages}{054111}
  (\bibinfo{year}{2016}).

\bibitem{s3dunning1989gaussian}
\bibinfo{author}{Dunning~Jr, T.~H.}
\newblock \bibinfo{title}{Gaussian basis sets for use in correlated molecular
  calculations. i. the atoms boron through neon and hydrogen}.
\newblock \emph{\bibinfo{journal}{J. Chem. Phys.}}
  \textbf{\bibinfo{volume}{90}}, \bibinfo{pages}{1007--1023}
  (\bibinfo{year}{1989}).

\bibitem{s4liu2006infinite}
\bibinfo{author}{Liu, W.} \& \bibinfo{author}{Peng, D.}
\newblock \bibinfo{title}{Infinite-order quasirelativistic density functional
  method based on the exact matrix quasirelativistic theory}.
\newblock \emph{\bibinfo{journal}{J. Chem. Phys.}}
  \textbf{\bibinfo{volume}{125}}, \bibinfo{pages}{044102}
  (\bibinfo{year}{2006}).

\bibitem{CAS1}
\bibinfo{author}{Werner, H.} \& \bibinfo{author}{Knowles, P.}
\newblock \bibinfo{title}{A second order multiconfiguration scf procedure with
  optimum convergence}.
\newblock \emph{\bibinfo{journal}{J. Chem. Phys.}}
  \textbf{\bibinfo{volume}{82}}, \bibinfo{pages}{5053--5063}
  (\bibinfo{year}{1985}).

\bibitem{CAS2}
\bibinfo{author}{Knowles, P.~J.} \& \bibinfo{author}{Werner, H.~L.}
\newblock \bibinfo{title}{An efficient second-order mc scf method for long
  configuration expansions}.
\newblock \emph{\bibinfo{journal}{Chem. Phys. Lett.}}
  \textbf{\bibinfo{volume}{115}}, \bibinfo{pages}{259--267}
  (\bibinfo{year}{1985}).

\bibitem{MRCI}
\bibinfo{author}{Werner, H.~J.} \& \bibinfo{author}{Knowles, P.~J.}
\newblock \bibinfo{title}{An efficient internally contracted
  multiconfiguration-reference configuration interaction method}.
\newblock \emph{\bibinfo{journal}{J. Chem. Phys.}}
  \textbf{\bibinfo{volume}{89}}, \bibinfo{pages}{5803--5814}
  (\bibinfo{year}{1988}).

\bibitem{MRCI2}
\bibinfo{author}{Shamasundar, K.~R.}, \bibinfo{author}{Knizia, G.} \&
  \bibinfo{author}{Werner, H.~L.}
\newblock \bibinfo{title}{A new internally contracted multi-reference
  configuration interaction method}.
\newblock \emph{\bibinfo{journal}{J. Chem. Phys.}}
  \textbf{\bibinfo{volume}{135}}, \bibinfo{pages}{054101}
  (\bibinfo{year}{2011}).

\bibitem{NEVPT2}
\bibinfo{author}{Angeli, C.}, \bibinfo{author}{Cimiraglia, R.} \&
  \bibinfo{author}{P, M.~J.}
\newblock \bibinfo{title}{n-electron valence state perturbation theory: A
  spinless formulation and an efficient implementation of the strongly
  contracted and of the partially contracted variants}.
\newblock \emph{\bibinfo{journal}{J. Chem. Phys.}}
  \textbf{\bibinfo{volume}{117}}, \bibinfo{pages}{9138--9153}
  (\bibinfo{year}{2002}).

\bibitem{SO}
\bibinfo{author}{Berning, A.}, \bibinfo{author}{Schweizer, M.},
  \bibinfo{author}{Werner, H.~J.}, \bibinfo{author}{Knowles, P.~J.} \&
  \bibinfo{author}{Palmieri, P.}
\newblock \bibinfo{title}{Spin-orbit matrix elements for internally contracted
  multireference configuration interaction wavefunctions}.
\newblock \emph{\bibinfo{journal}{Mol. Phys.}} \textbf{\bibinfo{volume}{98}},
  \bibinfo{pages}{1823--1833} (\bibinfo{year}{2000}).

\bibitem{antonov2002electronic1}
\bibinfo{author}{Antonov, V.}, \bibinfo{author}{Harmon, B.} \&
  \bibinfo{author}{Yaresko, A.}
\newblock \bibinfo{title}{Electronic structure of mixed-valence semiconductors
  in the lsda+ u approximation. ii. $\text{SmB}_6$ and $\text{YbB}_{12}$}.
\newblock \emph{\bibinfo{journal}{Phys. Rev. B}} \textbf{\bibinfo{volume}{66}},
  \bibinfo{pages}{165209} (\bibinfo{year}{2002}).

\bibitem{antonov2002electronic2}
\bibinfo{author}{Antonov, V.}, \bibinfo{author}{Harmon, B.} \&
  \bibinfo{author}{Yaresko, A.}
\newblock \bibinfo{title}{Electronic structure of mixed-valence semiconductors
  in the lsda+ u approximation. i. sm monochalcogenides}.
\newblock \emph{\bibinfo{journal}{Phys. Rev. B}} \textbf{\bibinfo{volume}{66}},
  \bibinfo{pages}{165208} (\bibinfo{year}{2002}).

\bibitem{vasp1}
\bibinfo{author}{Kresse, G.} \& \bibinfo{author}{Hafner, J.}
\newblock \bibinfo{title}{Ab initio molecular dynamics for liquid metals}.
\newblock \emph{\bibinfo{journal}{Phys. Rev. B}} \textbf{\bibinfo{volume}{47}},
  \bibinfo{pages}{558} (\bibinfo{year}{1993}).

\bibitem{vasp2}
\bibinfo{author}{Kresse, G.} \& \bibinfo{author}{Hafner, J.}
\newblock \bibinfo{title}{Ab initio molecular dynamics for liquid metals}.
\newblock \emph{\bibinfo{journal}{Phys. Rev. B}} \textbf{\bibinfo{volume}{47}},
  \bibinfo{pages}{558} (\bibinfo{year}{1993}).

\bibitem{vasp3}
\bibinfo{author}{Kresse, G.} \& \bibinfo{author}{Furthm{\"u}ller, J.}
\newblock \bibinfo{title}{Efficiency of ab-initio total energy calculations for
  metals and semiconductors using a plane-wave basis set}.
\newblock \emph{\bibinfo{journal}{Comput. Mater. Sci}}
  \textbf{\bibinfo{volume}{6}}, \bibinfo{pages}{15--50} (\bibinfo{year}{1996}).

\bibitem{vasp4}
\bibinfo{author}{Kresse, G.} \& \bibinfo{author}{Furthm{\"u}ller, J.}
\newblock \bibinfo{title}{Efficient iterative schemes for ab initio
  total-energy calculations using a plane-wave basis set}.
\newblock \emph{\bibinfo{journal}{Phys. Rev. B}} \textbf{\bibinfo{volume}{54}},
  \bibinfo{pages}{11169} (\bibinfo{year}{1996}).

\bibitem{CCSDT}
\bibinfo{author}{Watts, J.~D.}, \bibinfo{author}{Gauss, J.} \&
  \bibinfo{author}{Bartlett, R.~J.}
\newblock \bibinfo{title}{Coupled‐cluster methods with noniterative triple
  excitations for restricted open‐shell hartree–fock and other general
  single determinant reference functions. energies and analytical gradients}.
\newblock \emph{\bibinfo{journal}{J. Chem. Phys.}}
  \textbf{\bibinfo{volume}{98}}, \bibinfo{pages}{8718--8733}
  (\bibinfo{year}{1993}).

\bibitem{G09}
\bibinfo{author}{Frisch, M.~J.} \emph{et~al.}
\newblock \bibinfo{title}{Gaussian 09 {R}evision {D}.01}
  (\bibinfo{year}{2009}).
\newblock \bibinfo{note}{Gaussian Inc. Wallingford CT}.

\bibitem{s5sakai2007effect}
\bibinfo{author}{Sakai, T.}, \bibinfo{author}{Oomi, G.},
  \bibinfo{author}{Uwatoko, Y.} \& \bibinfo{author}{Kunii, S.}
\newblock \bibinfo{title}{Effect of pressure on the metamagnetic transition of
  $\text{DyB}_6$ single crystal}.
\newblock \emph{\bibinfo{journal}{J. Magn. Magn. Mater.}}
  \textbf{\bibinfo{volume}{310}}, \bibinfo{pages}{1732--1734}
  (\bibinfo{year}{2007}).

\bibitem{s6dorenbos2003systematic}
\bibinfo{author}{Dorenbos, P.}
\newblock \bibinfo{title}{Systematic behaviour in trivalent lanthanide charge
  transfer energies}.
\newblock \emph{\bibinfo{journal}{J. Phys. Condens. Matter}}
  \textbf{\bibinfo{volume}{15}}, \bibinfo{pages}{8417} (\bibinfo{year}{2003}).

\bibitem{7kim2014topological}
\bibinfo{author}{Kim, D.-J.}, \bibinfo{author}{Xia, J.} \&
  \bibinfo{author}{Fisk, Z.}
\newblock \bibinfo{title}{Topological surface state in the kondo insulator
  samarium hexaboride}.
\newblock \emph{\bibinfo{journal}{Nat. Mater.}} \textbf{\bibinfo{volume}{13}},
  \bibinfo{pages}{466} (\bibinfo{year}{2014}).

\bibitem{s8goodrich1998fermi}
\bibinfo{author}{Goodrich, R.} \emph{et~al.}
\newblock \bibinfo{title}{Fermi surface of ferromagnetic $\text{EuB}_6$}.
\newblock \emph{\bibinfo{journal}{Phys. Rev. B}} \textbf{\bibinfo{volume}{58}},
  \bibinfo{pages}{14896} (\bibinfo{year}{1998}).

\bibitem{s9weill1979electrical}
\bibinfo{author}{Weill, G.}, \bibinfo{author}{Gurin, V.} \emph{et~al.}
\newblock \bibinfo{title}{Electrical transport properties of $\text{EuB}_6$
  under pressure}.
\newblock \emph{\bibinfo{journal}{Phys. Status Solidi A}}
  \textbf{\bibinfo{volume}{53}} (\bibinfo{year}{1979}).

\bibitem{s10fujita1980specific}
\bibinfo{author}{Fujita, T.}, \bibinfo{author}{Suzuki, M.} \&
  \bibinfo{author}{Ishikawa, Y.}
\newblock \bibinfo{title}{Specific heat of $\text{EuB}_6$}.
\newblock \emph{\bibinfo{journal}{Solid State Commun.}}
  \textbf{\bibinfo{volume}{33}}, \bibinfo{pages}{947--950}
  (\bibinfo{year}{1980}).

\bibitem{s11hidaka2009specific}
\bibinfo{author}{Hidaka, H.}, \bibinfo{author}{Ikeda, Y.},
  \bibinfo{author}{Kawasaki, I.}, \bibinfo{author}{Yanagisawa, T.} \&
  \bibinfo{author}{Amitsuka, H.}
\newblock \bibinfo{title}{Specific heat of $\text{EuIn}_2\text{P}_2$ at high
  magnetic fields}.
\newblock \emph{\bibinfo{journal}{Physica B: Condensed Matter}}
  \textbf{\bibinfo{volume}{404}}, \bibinfo{pages}{3005--3007}
  (\bibinfo{year}{2009}).

\bibitem{s12nozaki1980magnetic}
\bibinfo{author}{Nozaki, H.}, \bibinfo{author}{Tanaka, T.} \&
  \bibinfo{author}{Ishizawa, Y.}
\newblock \bibinfo{title}{Magnetic behaviour and structure change of
  $\text{GdB}_6$ single crystals at low temperatures}.
\newblock \emph{\bibinfo{journal}{Journal of Physics C: Solid State Physics}}
  \textbf{\bibinfo{volume}{13}}, \bibinfo{pages}{2751} (\bibinfo{year}{1980}).

\bibitem{s13kunii1985electronic}
\bibinfo{author}{Kunii, S.} \emph{et~al.}
\newblock \bibinfo{title}{Electronic and magnetic properties of
  $\text{GdB}_6$}.
\newblock \emph{\bibinfo{journal}{J. Magn. Magn. Mater.}}
  \textbf{\bibinfo{volume}{52}}, \bibinfo{pages}{275--278}
  (\bibinfo{year}{1985}).

\bibitem{s14grechnev2012pressure}
\bibinfo{author}{Grechnev, G.}, \bibinfo{author}{Logosha, A.},
  \bibinfo{author}{Panfilov, A.} \& \bibinfo{author}{Shitsevalova, N.~Y.}
\newblock \bibinfo{title}{Pressure effects on magnetic properties and
  electronic structure of $\text{EuB}_6$ and $\text{GdB}_6$}.
\newblock \emph{\bibinfo{journal}{Journal of Alloys and Compounds}}
  \textbf{\bibinfo{volume}{511}}, \bibinfo{pages}{5--8} (\bibinfo{year}{2012}).

\bibitem{s15mccarthy1980magnetic}
\bibinfo{author}{McCarthy, C.} \& \bibinfo{author}{Tompson, C.}
\newblock \bibinfo{title}{Magnetic structure of $\text{NdB}_6$}.
\newblock \emph{\bibinfo{journal}{Journal of Physics and Chemistry of Solids}}
  \textbf{\bibinfo{volume}{41}}, \bibinfo{pages}{1319--1321}
  (\bibinfo{year}{1980}).

\bibitem{s16goodrich2006fermi}
\bibinfo{author}{Goodrich, R.}, \bibinfo{author}{Harrison, N.} \&
  \bibinfo{author}{Fisk, Z.}
\newblock \bibinfo{title}{Fermi surface changes across the n{\'e}el phase
  boundary of $\text{NdB}_6$}.
\newblock \emph{\bibinfo{journal}{Phys. Rev. Lett.}}
  \textbf{\bibinfo{volume}{97}}, \bibinfo{pages}{146404}
  (\bibinfo{year}{2006}).

\bibitem{s17lazukov2016thermal}
\bibinfo{author}{Lazukov, V.}, \bibinfo{author}{Alekseev, P.},
  \bibinfo{author}{Shitsevalova, N.~Y.} \& \bibinfo{author}{Philippov, V.}
\newblock \bibinfo{title}{Thermal evolution of magnetic-excitation spectrum of
  $\text{PrB}_6$}.
\newblock \emph{\bibinfo{journal}{The Physics of Metals and Metallography}}
  \textbf{\bibinfo{volume}{117}}, \bibinfo{pages}{460--465}
  (\bibinfo{year}{2016}).

\bibitem{s18kuromaru2002multipolar}
\bibinfo{author}{Kuromaru, T.}, \bibinfo{author}{Kusunose, H.} \&
  \bibinfo{author}{Kuramoto, Y.}
\newblock \bibinfo{title}{Multipolar ordering in $\text{PrB}_6$}.
\newblock \emph{\bibinfo{journal}{J. Phys. Soc. Jpn.}}
  \textbf{\bibinfo{volume}{71}}, \bibinfo{pages}{130--132}
  (\bibinfo{year}{2002}).

\bibitem{s19sera2004crystal}
\bibinfo{author}{Sera, M.}, \bibinfo{author}{Kim, M.-S.}, \bibinfo{author}{Tou,
  H.} \& \bibinfo{author}{Kunii, S.}
\newblock \bibinfo{title}{Crystal structure and magnetic anisotropy in the
  magnetic ordered phases of $\text{PrB}_6$}.
\newblock \emph{\bibinfo{journal}{J. Phys. Soc. Jpn.}}
  \textbf{\bibinfo{volume}{73}}, \bibinfo{pages}{3422--3428}
  (\bibinfo{year}{2004}).

\bibitem{s20deng2013plutonium}
\bibinfo{author}{Deng, X.}, \bibinfo{author}{Haule, K.} \&
  \bibinfo{author}{Kotliar, G.}
\newblock \bibinfo{title}{Plutonium hexaboride is a correlated topological
  insulator}.
\newblock \emph{\bibinfo{journal}{Phys. Rev. Lett.}}
  \textbf{\bibinfo{volume}{111}}, \bibinfo{pages}{176404}
  (\bibinfo{year}{2013}).

\bibitem{s21xu2014direct}
\bibinfo{author}{Xu, N.} \emph{et~al.}
\newblock \bibinfo{title}{Direct observation of the spin texture in
  $\text{SmB}_6$ as evidence of the topological kondo insulator}.
\newblock \emph{\bibinfo{journal}{Nat. Commun.}} \textbf{\bibinfo{volume}{5}},
  \bibinfo{pages}{4566} (\bibinfo{year}{2014}).

\bibitem{s22zhou2015pressure}
\bibinfo{author}{Zhou, Y.} \emph{et~al.}
\newblock \bibinfo{title}{Pressure-induced quantum phase transitions in a
  $\text{YbB}_6$ single crystal}.
\newblock \emph{\bibinfo{journal}{Phys. Rev. B}} \textbf{\bibinfo{volume}{92}},
  \bibinfo{pages}{241118} (\bibinfo{year}{2015}).

\bibitem{s23akang2016electronic}
\bibinfo{author}{Kang, C.-J.} \emph{et~al.}
\newblock \bibinfo{title}{Electronic structure of $\text{YbB}_6$: Is it a
  topological insulator or not?}
\newblock \emph{\bibinfo{journal}{Phys. Rev. Lett.}}
  \textbf{\bibinfo{volume}{116}}, \bibinfo{pages}{116401}
  (\bibinfo{year}{2016}).

\bibitem{s24abxiang2018quantum}
\bibinfo{author}{Xiang, Z.} \emph{et~al.}
\newblock \bibinfo{title}{Quantum oscillations of electrical resistivity in an
  insulator}.
\newblock \emph{\bibinfo{journal}{Science}} \textbf{\bibinfo{volume}{362}},
  \bibinfo{pages}{65--69} (\bibinfo{year}{2018}).

\bibitem{s24balekseev2015high}
\bibinfo{author}{Alekseev, P.~A.}
\newblock \bibinfo{title}{High borides: determining the features and details of
  lattice dynamics from neutron spectroscopy}.
\newblock \emph{\bibinfo{journal}{Phys.-Uspekhi}}
  \textbf{\bibinfo{volume}{58}}, \bibinfo{pages}{330} (\bibinfo{year}{2015}).

\bibitem{s25kanai2015evidence}
\bibinfo{author}{Kanai, Y.} \emph{et~al.}
\newblock \bibinfo{title}{Evidence for $\gamma$8 ground-state symmetry of cubic
  $\text{YbB}_{12}$ probed by linear dichroism in core-level photoemission}.
\newblock \emph{\bibinfo{journal}{J. Phys. Soc. Jpn.}}
  \textbf{\bibinfo{volume}{84}}, \bibinfo{pages}{073705}
  (\bibinfo{year}{2015}).

\bibitem{s26HollasSpectro}
\bibinfo{author}{Hollas, M.~J.}
\newblock \emph{\bibinfo{title}{Modern Spectroscopy}}
  (\bibinfo{publisher}{Wiley}, \bibinfo{year}{1996}), \bibinfo{edition}{3} edn.

\bibitem{s27kim2012temperature}
\bibinfo{author}{Kim, Y.} \emph{et~al.}
\newblock \bibinfo{title}{Temperature dependence of raman-active optical
  phonons in $\text{Bi}_2\text{Se}_3$ and $\text{Sb}_2\text{Te}_3$}.
\newblock \emph{\bibinfo{journal}{Appl. Phys. Lett.}}
  \textbf{\bibinfo{volume}{100}}, \bibinfo{pages}{071907}
  (\bibinfo{year}{2012}).

\bibitem{s28zhang2015observation}
\bibinfo{author}{Zhang, W.-L.} \emph{et~al.}
\newblock \bibinfo{title}{Observation of a raman-active phonon with fano line
  shape in the quasi-one-dimensional superconductor
  $\text{K}_2\text{Cr}_3\text{As}_3$}.
\newblock \emph{\bibinfo{journal}{Phys. Rev. B}} \textbf{\bibinfo{volume}{92}},
  \bibinfo{pages}{060502} (\bibinfo{year}{2015}).

\bibitem{s29fano1961effects}
\bibinfo{author}{Fano, U.}
\newblock \bibinfo{title}{Effects of configuration interaction on intensities
  and phase shifts}.
\newblock \emph{\bibinfo{journal}{Phys. Rev.}} \textbf{\bibinfo{volume}{124}},
  \bibinfo{pages}{1866} (\bibinfo{year}{1961}).

\end{thebibliography}

\end{document}


\preprint{}

\title{Supplementary Information: Dynamical Bonding Driving Mixed Valency in a Metal Boride}

\author{Paul J Robinson}
    \thanks{Current Address: Department of Chemistry, Columbia University, New York, New York 10027, USA}
    \affiliation{Department of Chemistry and Biochemistry, University of California Los Angeles, Los Angeles, California 90095, USA}
\author{Julen Munarriz}
    \affiliation{Department of Chemistry and Biochemistry, University of California Los Angeles, Los Angeles, California 90095, USA}
\author{Michael E. Valentine}
    \affiliation{Institute for Quantum Matter, Department of Physics and Astronomy, The Johns Hopkins University, Baltimore, Maryland 21218, USA}
\author{Austin Granmoe} 
    \affiliation{Institute for Quantum Matter, Department of Physics and Astronomy, The Johns Hopkins University, Baltimore, Maryland 21218, USA}
\author{Natalia Drichko}
    \email{drichko@jhu.edu, mcqueen@jhu.edu, ana@chem.ucla.edu }
    \affiliation{Institute for Quantum Matter, Department of Physics and Astronomy, The Johns Hopkins University, Baltimore, Maryland 21218, USA}
\author{Juan R. Chamorro}
    \affiliation{Institute for Quantum Matter, Department of Physics and Astronomy, The Johns Hopkins University, Baltimore, Maryland 21218, USA}
    \affiliation{Department of Chemistry, The Johns Hopkins University, Baltimore, Maryland 21218, USA}
\author{Priscila F. Rosa}
    \affiliation{Los Alamos National Laboratory, Los Alamos, NM 87545, USA}
\author{Tyrel M. McQueen}
    \email{drichko@jhu.edu, mcqueen@jhu.edu, ana@chem.ucla.edu }
    \affiliation{Institute for Quantum Matter, Department of Physics and Astronomy, The Johns Hopkins University, Baltimore, Maryland 21218, USA}
    \affiliation{Department of Chemistry, The Johns Hopkins University, Baltimore, Maryland 21218, USA}
    \affiliation{Department of Materials Science and Engineering, The Johns Hopkins University, Baltimore, Maryland 21218, USA}
\author{Anastassia N. Alexandrova}
    \email{drichko@jhu.edu, mcqueen@jhu.edu, ana@chem.ucla.edu }
    \affiliation{Department of Chemistry and Biochemistry, University of California Los Angeles, Los Angeles, California 90095, USA}
    \affiliation{California NanoSystems Institute, Los Angeles, California 90095, USA}

\date{\today}

\maketitle


\setcounter{section}{0}
\setcounter{equation}{0}
\setcounter{figure}{0}
\setcounter{table}{0}
\setcounter{page}{1}
\makeatletter
\renewcommand{\thepage}{S\arabic{page}}
\renewcommand{\thesection}{S\arabic{section}}
\renewcommand{\theequation}{S\arabic{equation}}
\renewcommand{\thefigure}{S\arabic{figure}}
\tableofcontents
\newpage
\section{Plane Wave DFT Calculations}

\begin{figure}[h!]
   \includegraphics[width=0.95\linewidth]{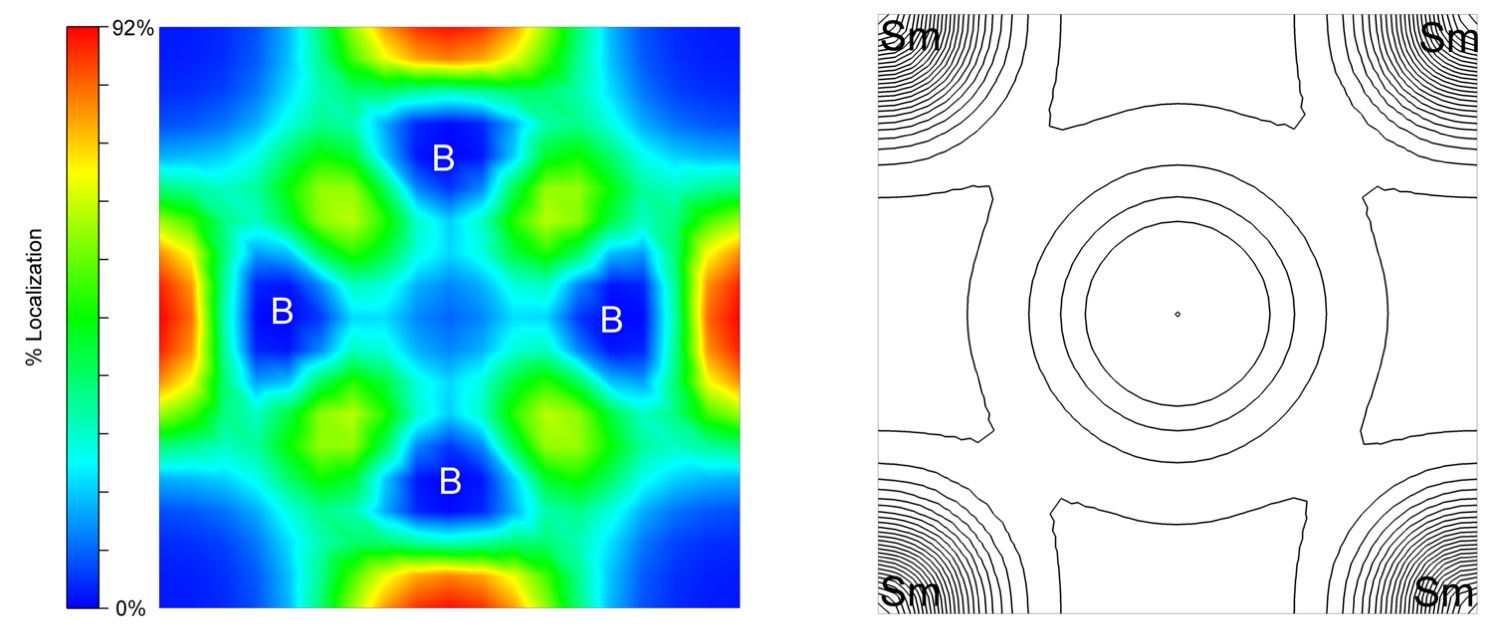}
   \caption{Left: The Electron Localization Function (ELF) of a crude DFT+U calculation. The electrons are significantly more localized between unit cells in the B$_2$ units than in the octahedron. Right: The charge density of the same calculation demonstrating a potential bonding interaction between a B$_2$ and the Sm.}
\end{figure}

\begin{figure}[p!]
   \includegraphics[width=0.95\linewidth]{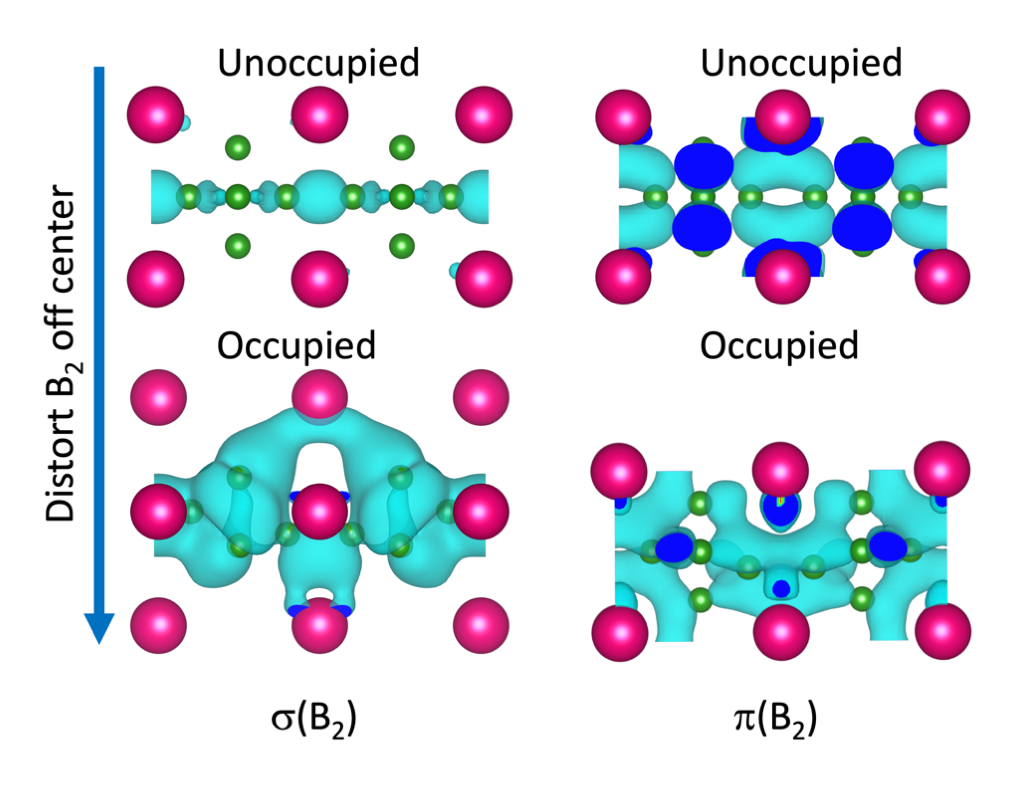}
   \caption{Artificial deformation displacing B$_2$ toward one of the Sm ions by 20 \% of the equilibrium Sm-B$_2$ distance. At the gamma point the B$_2$ $\sigma(p_z)$ state goes from virtual to valence in agreement with our cluster model.}
\end{figure}

\clearpage

\section{Computational Methods}
All the multi-reference calculations were performed with Molpro.\cite{s1werner2012molpro}
The cc-pvDz-DK3 \cite{s2lu2016correlation} and cc-pvDz \cite{s3dunning1989gaussian} basis sets were used for Sm and B, respectively. We selected the sixth-order Douglass-Kroll-Hess Hamiltonian \cite{s4liu2006infinite} 
to account for significant scalar relativistic effects beyond the second order present in Sm systems. 
Motivation for our choices of basis set and order of scalar relativistic correction 
can be found in a previous study of the $\text{SmB}_6^-$ cluster (Ref 33, main text). 
The potential energy surface of SmB$_2^+$ was modeled with 
Multi-reference Configuration Interactions (MRCI) starting from the 
8-SA-CASSCF(14o,9e) wave-function, as implemented in the 
Molpro suite.\cite{CAS1,CAS2,MRCI,MRCI2} 
For each point in the energy surface, we considered the lowest doublet, 
quartet, sextet and octet A$_1$, A$_2$, B$_1$ and B$_2$ states. 
This data is provided in Tables SI-SV. 
As a sanity check, we compared the MRCI energy in the two minima with that provided by 
SO(8-SA-CASSCF(14o,9e)/NEVPT2).\cite{NEVPT2} 
Spin-orbit effects were included via the state interacting approach,\cite{SO} 
that is, SO coupling was treated as a perturbation using unperturbed (spin-free) 
zeroth-order wave-functions. 
In particular, we considered the eight lowest-in-energy 
A$_1$, A$_2$, B$_1$ and B$_2$ quartet and sextet states in both minima. 
The results are in agreement with the MRCI ones, 
the energy difference between both minima being 10.3 meV. The optimized geometrical parameters are also very similar: for the first minimum, the Sm--B and B--B distances are 2.14 {\AA} and 1.61 {\AA} with both methods. For the second minimum, the Sm--B distance at the MRCI and NEVPT2 levels of theory are 2.25 {\AA} and 2.27 {\AA} respectively, and, in both cases, the B--B distance is 1.55 {\AA}. The optimized B$_2$ length was computed at each Sm-B$_2$ bond
length by fitting a quadratic function to several points sampled by the minimum.
The $\left[\text{B}_2\text{SmB}_2\right]^{2+}$ cluster was modeled in several 
different charges and states with a CASSCF(19o,12e)+DKH6. 
Because of the high symmetry and many accessible electronic configurations, 
we found it necessary even for our qualitative understanding to use an active space of this size.

It is important to confirm that our selected three-atom cluster does not ignore important interactions in the solid. Namely, does the addition of more than one B$_2$ unit change the bonding. 
To check this, we model the $\left[\text{B}_2\text{--Sm--B}_2\right]^{2+}$ cluster’s $^9\text{A}_2$ state. Optimized, we see that Sm bonds covalently with one of the B$_2$ units and is nearly non-interacting with the other as evidenced by the Sm-B$_2$bond lengths of 2.38 {\AA} and 3.09 {\AA} respectively. 
Considering that in the solid the Sm is always found in at least a +2 state we can understand this result. A Sm coordinated to 3 B$_2$ units (SmB$_6$) will have donated electrons to two of its three adjacent B$_2$ units making them non-interacting. Knowing this, we return to our smaller, more approachable, model to understand the intricacies of Sm-B$_2$ bonding while remaining confident that only one B$_2$ interacts with a single Sm at a time.

DFT+U calculations have previously been used to successfully describe SmB$_6$ and a range of mixed valent semiconductors \cite{antonov2002electronic1, antonov2002electronic2}.
So, for our exploratory plane wave DFT calculations, 
the SmB$_6$ solid was treated using VASP 5.2, PBE-D2+U (U=7.0 eV for Sm) 
and the PAW PBE plane wave basis sets.
\cite{vasp1, vasp2, vasp3, vasp4}
For examining the differences between the homotopic samarium atoms, 
a 1x1x2 super cell was calculated with a k-mesh of 3x1x3 and an energy cutoff of 318 eV.

The vertical electron detachment energies (VDE) of $\text{B}_2^{-}$ were computed at the ROCCSD(T) level of theory,\cite{CCSDT} using the cc-pVTZ basis set \cite{s3dunning1989gaussian}, as implemented in the Gaussian09 package.\cite{G09} Notice that the geometry optimization of $\text{B}_2^{-}$ was performed using the same method.
\newpage

\section{$\text{SmB}_2^{+}$ Cluster Results}

\begin{table} [h!]
   \caption{
   Energy (relative to the global minimum) and optimized B-B distance and ground state symmetry as a function of the Sm-B$_2$ distance.
   }
 \begin{tabular}{ c | c | c | c} 
 \hline
  d(Sm-B$_2$) &	d(B-B)	& E(a.u.) &	$\Delta E(eV)$ \\  
 \hline
 2.000	&   1.611	&  -10472.54024006   &	 0.239 \\ 
 2.025	&   1.609	&  -10472.54322079   &   0.158 \\
 2.050	&   1.607   &  -10472.54553611   &   0.095 \\
 2.075	&   1.607   &  -10472.54722990   &	 0.049 \\
 2.100	&   1.607	&  -10472.54834499   & 	 0.019 \\ 
 2.125  &	1.606   &  -10472.54892306   &	 0.003 \\
 2.143  &   1.605   &  -10472.54902924   &	 0.000 \\
 2.150  &	1.605   &  -10472.54900424   &	 0.001 \\
 2.175  &	1.604   &  -10472.54862636   &	 0.011 \\
 2.200  &	1.605   &  -10472.54782744   &	 0.033 \\
 2.225  &	1.545   &  -10472.54840617   &	 0.017 \\
 2.250  &	1.546   &  -10472.54885656   &	 0.005 \\
 2.254  &	1.546   &  -10472.54888956   &	 0.004 \\
 2.275  &	1.556   &  -10472.54862204   &	 0.007 \\
 2.300  &	1.564   &  -10472.54839098   &	 0.017 \\
 2.325  &	1.574   &  -10472.54747415   &	 0.042 \\
 2.350  &	1.582   &  -10472.54622053   &	 0.076
\end{tabular}
\end{table}

\clearpage

\begin{table} [h!]
   \caption{
   Doublets energy (a.u.) as a function of the Sm-B$_2$ distance (\AA).
   }
 \begin{tabular}{ c | c | c | c | c} 
 \hline
  d(Sm-B$_2$) &	$^2$A$_1$	& $^2$B$_1$ & $^2$B$_2$ & $^2$A$_2$ \\  
 \hline
2.000	&		-10472.5274683	&		-10472.5289491	&		-10472.5307758	&		-10472.5326172	\\
2.025	&		-10472.5307351	&		-10472.5319490	&		-10472.5338625	&		-10472.5358904	\\
2.050	&		-10472.5333094	&		-10472.5342771	&		-10472.5362731	&		-10472.5384689	\\
2.075	&		-10472.5352371	&		-10472.5359739	&		-10472.5380517	&		-10472.5403998	\\
2.100	&		-10472.5365630	&		-10472.5370872	&		-10472.5392433	&		-10472.5417284	\\
2.125	&		-10472.5373327	&		-10472.5376634	&		-10472.5398921	&		-10472.5425004	\\
2.143	&		-10472.5375670	&		-10472.5377664	&		-10472.5400439	&		-10472.5427346	\\
2.150	&		-10472.5375875	&		-10472.5377395	&		-10472.5400350	&		-10472.5427551	\\
2.175	&		-10472.5373664	&		-10472.5373523	&		-10472.5397106	&		-10472.5425330	\\
2.200	&		-10472.5367079	&		-10472.5365361	&		-10472.5389525	&		-10472.5418708	\\
2.225	&		-10472.5214622	&		-10472.5258276	&		-10472.5300100	&		-10472.5260687	\\
2.250	&		-10472.5220984	&		-10472.5266024	&		-10472.5308624	&		-10472.5266024	\\
2.254	&		-10472.5221929	&		-10472.5267283	&		-10472.5309716	&		-10472.5266853	\\
2.275	&		-10472.5234747	&		-10472.5279938	&		-10472.5322566	&		-10472.5278816	\\
2.300	&		-10472.5233718	&		-10472.5278417	&		-10472.5321630	&		-10472.5277463	\\
2.325	&		-10472.5234768	&		-10472.5278726	&		-10472.5322310	&		-10472.5278000	\\
2.350	&		-10472.5231505	&		-10472.5274679	&		-10472.5318627	&		-10472.5274316	\\
\end{tabular}
\end{table}

\clearpage

\begin{table}[h!]
   \caption{
   Quartets energy (a.u.) as a function of the Sm-B$_2$ distance (\AA).
   }
 \begin{tabular}{ c | c | c | c | c} 
 \hline
  d(Sm-B$_2$) &	$^4$A$_1$	& $^4$B$_1$ & $^4$B$_2$ & $^4$A$_2$ \\  
 \hline
2.000	&		-10472.5306487	&		-10472.5304853	&		-10472.5327377	&		-10472.5353849	\\
2.025	&		-10472.5337588	&		-10472.5334417	&		-10472.5357747	&		-10472.5385465	\\
2.050	&		-10472.5361909	&		-10472.5357317	&		-10472.5381397	&		-10472.5410231	\\
2.075	&		-10472.5379895	&		-10472.5373964	&		-10472.5398782	&		-10472.5428615	\\
2.100	&		-10472.5391987	&		-10472.5384813	&		-10472.5410332	&		-10472.5441060	\\
2.125	&		-10472.5398630	&		-10472.5390318	&		-10472.5416480	&		-10472.5448012	\\
2.143	&		-10472.5400269	&		-10472.5391183	&		-10472.5417771	&		-10472.5449840	\\
2.150	&		-10472.5400224	&		-10472.5390860	&		-10472.5417603	&		-10472.5449862	\\
2.175	&		-10472.5397151	&		-10472.5386811	&		-10472.5414099	&		-10472.5447014	\\
2.200	&		-10472.5389787	&		-10472.5378535	&		-10472.5406320	&		-10472.5439835	\\
2.225	&		-10472.5432000	&		-10472.5418074	&		-10472.5454108	&		-10472.5484062	\\
2.250	&		-10472.5436379	&		-10472.5422287	&		-10472.5458329	&		-10472.5488566	\\
2.254	&		-10472.5436694	&		-10472.5422571	&		-10472.5458613	&		-10472.5488896	\\
2.275	&		-10472.5434452	&		-10472.5419941	&		-10472.5455747	&		-10472.5486220	\\
2.300	&		-10472.5431904	&		-10472.5417271	&		-10472.5453177	&		-10472.5483910	\\
2.325	&		-10472.5422878	&		-10472.5407954	&		-10472.5443795	&		-10472.5474742	\\
2.350	&		-10472.5410436	&		-10472.5395259	&		-10472.5431065	&		-10472.5462205	\\
\end{tabular}
\end{table}

\clearpage

\begin{table} [h!]
   \caption{
   Sextets energy (a.u.) as a function of the Sm-B$_2$ distance (\AA).
   }
 \begin{tabular}{ c | c | c | c | c} 
 \hline
  d(Sm-B$_2$) &	$^6$A$_1$	& $^6$B$_1$ & $^6$B$_2$ & $^6$A$_2$ \\  
 \hline
2.000	&		-10472.535524	&		-10472.533180	&		-10472.536750	&		-10472.540240	\\
2.025	&		-10472.538439	&		-10472.536045	&		-10472.539678	&		-10472.543221	\\
2.050	&		-10472.540696	&		-10472.538254	&		-10472.541946	&		-10472.545536	\\
2.075	&		-10472.542339	&		-10472.539851	&		-10472.543597	&		-10472.547230	\\
2.100	&		-10472.543408	&		-10472.540876	&		-10472.544674	&		-10472.548345	\\
2.125	&		-10472.543948	&		-10472.541372	&		-10472.545216	&		-10472.548923	\\
2.143	&		-10472.544029	&		-10472.541423	&		-10472.545298	&		-10472.549029	\\
2.150	&		-10472.543996	&		-10472.541378	&		-10472.545265	&		-10472.549004	\\
2.175	&		-10472.543590	&		-10472.540932	&		-10472.544855	&		-10472.548626	\\
2.200	&		-10472.542768	&		-10472.540071	&		-10472.544028	&		-10472.547827	\\
2.225	&		-10472.525227	&		-10472.527866	&		-10472.532359	&		-10472.528733	\\
2.250	&		-10472.526029	&		-10472.528700	&		-10472.533208	&		-10472.529551	\\
2.254	&		-10472.526129	&		-10472.528805	&		-10472.533316	&		-10472.529654	\\
2.275	&		-10472.527675	&		-10472.530395	&		-10472.534864	&		-10472.531183	\\
2.300	&		-10472.527383	&		-10472.530127	&		-10472.534628	&		-10472.530921	\\
2.325	&		-10472.527461	&		-10472.530240	&		-10472.534735	&		-10472.531008	\\
\end{tabular}
\end{table}

\clearpage

\begin{table} [h!]
   \caption{
   Octets energy (a.u.) as a function of the Sm-B$_2$ distance (\AA).
   }
 \begin{tabular}{ c | c | c | c | c  | c } 
 \hline
  d(Sm-B$_2$) &	$^8$A$_1$	& $^8$B$_1$ & $^8$B$_2$ & $^8$A$_2$  \\  
 \hline
2.000	&		-10472.4787079	&		-10472.4742590	&		-10472.4716170	&		-10472.4747142	\\
2.025	&		-10472.4821786	&		-10472.4778417	&		-10472.4743014	&		-10472.4776341	\\
2.050	&		-10472.4755980	&		-10472.4797080	&		-10472.4763203	&		-10472.4716739	\\
2.075	&		-10472.4797832	&		-10472.4838644	&		-10472.4805279	&		-10472.4759927	\\
2.100	&		-10472.4834860	&		-10472.4875411	&		-10472.4842580	&		-10472.4798097	\\
2.125	&		-10472.4884675	&		-10472.4936603	&		-10472.4928039	&		-10472.4873702	\\
2.143	&		-10472.4907047	&		-10472.4947112	&		-10472.4918412	&		-10472.4877357	\\
2.150	&		-10472.4914047	&		-10472.4954111	&		-10472.4925493	&		-10472.4884405	\\
2.175	&		-10472.4937515	&		-10472.4977586	&		-10472.4949246	&		-10472.4908029	\\
2.200	&		-10472.4958532	&		-10472.4998621	&		-10472.4970518	&		-10472.4929168	\\
2.225	&		-10472.4923500	&		-10472.4964091	&		-10472.4936380	&		-10472.4894184	\\
2.250	&		-10472.4939293	&		-10472.4979944	&		-10472.4952362	&		-10472.4909991	\\
2.254	&		-10472.4941736	&		-10472.4982398	&		-10472.4954834	&		-10472.4912434	\\
2.275	&		-10472.4975847	&		-10472.5016405	&		-10472.4988891	&		-10472.4946572	\\
2.300	&		-10472.4979389	&		-10472.5020100	&		-10472.4992693	&		-10472.4950072	\\
2.325	&		-10472.4994881	&		-10472.5035648	&		-10472.5008322	&		-10472.4965549	\\
2.350	&		-10472.5021352	&		-10472.5080162	&		-10472.5079530	&		-10472.5019627	\\
\end{tabular}
\end{table}

\clearpage

\begin{table} [h!]
   \caption{
   Natural populations for the \textit{f}-orbitals in the active space for the different symmetry and spin states of the $\text{SmB}_2^{2+}$ cluster. Notice that this cluster is the result of the vertical detachment of one electron from the first minimum in $\text{SmB}_2^{+}$ potential energy surface. The relative energy of the $\text{SmB}_2^{2+}$ states for this geometry is also provided
   }
 \begin{tabular}{ c | c | c | c | c | c | c | c } 
 \hline
 min$_1$ &	$^2$a$_1$ & $^2$b$_1$ & $^3$b$_1$ & $^2$b$_2$ & $^4$b$_2$ & $^1$a$_2$ & $\Delta E (eV)$ \\
\hline
$\text{SmB}_2^{+}$	&		0.57	&		0.59	&		0.43	&		0.61	&		0.05	&		0.74  & $--$  \\
\hline
$^3$A$_1$ 	&		0.50	&		0.01	&		0.49	&		1.00	&		0.49	&		0.51 & 0.385 \\
$^3$B$_1$ &		0.09	&		1.09	&		0.01	&		1.00	&		0.01	&		1.00 & 0.275 \\
$^3$B$_2$ 	&		0.43	&		0.90	&		0.56	&		0.43	&		0.04	&		0.57  & 0.171 \\
$^3$A$_2$ 	&		0.46	&		0.96	&		0.55	&		0.45	&		0.03	&		0.54 & 0.298 \\
$^5$A$_1$	&		0.47	&		0.01	&		0.52	&		1.00	&		0.53	&		0.47 & 0.308 \\
$^5$B$_1$	&		0.02	&		0.92	&		0.01	&		1.00	&		0.01	&		1.00  & 0.215 \\
$^5$B$_2$	&		0.44	&		0.93	&		0.56	&		0.43	&		0.04	&		0.56 & 0.102 \\
$^5$A$_2$	&		0.45	&		0.96	&		0.56	&		0.44	&		0.04	&		0.55 & 0.218  \\
$^7$A$_1$	&		0.42	&		0.01	&		0.59	&		1.00	&		0.59	&		0.42 & 0.196 \\
$^7$B$_1$	&		0.01	&		1.00	&		0.02	&		1.00	&		0.02	&		1.00 & 0.121 \\
$^7$B$_2$	&		0.44	&		0.96	&		0.56	&		0.44	&		0.04	&		0.56 & 0.000 \\
$^7$A$_2$	&	0.94	&		1.00	&		0.06	&		0.01	&		0.05	&		0.94 & 0.108 \\

\end{tabular}
\end{table}

\clearpage

\begin{table} [h!]
   \caption{
   Natural populations for the \textit{f}-orbitals in the active space for the different symmetry and spin states of the $\text{SmB}_2^{2+}$ cluster. Notice that this cluster is the result of the vertical detachment of one electron from the second minimum in $\text{SmB}_2^{+}$ potential energy surface. The relative energy of the $\text{SmB}_2^{2+}$ states for this geometry is also provided
   }
 \begin{tabular}{ c | c | c | c | c | c | c | c } 
 \hline
 min$_2$ &	$^2$a$_1$ & $^2$b$_1$ & $^3$b$_1$ & $^2$b$_2$ & $^4$b$_2$ & $^1$a$_2$ & $\Delta E (eV)$ \\
\hline
$\text{SmB}_2^{+}$	&		0.59	&		0.59	&		0.44	&		0.61	&		0.05	&		0.74  & $--$  \\
\hline
$^3$A$_1$ &		0.94	&		0.06	&		0.02	&		1.00	&		0.06	&		0.94	&		0.339    \\
$^3$B$_1$ &		0.01	&		0.85	&		0.02	&		1.00	&		0.01	&		0.99	&		0.172   \\
$^3$B$_2$ &		0.42	&		0.89	&		0.57	&		0.42	&		0.04	&		0.58	&		0.070	 \\
$^3$A$_2$ &		0.38	&		0.98	&		0.59	&		0.62	&		0.00	&		0.38	&		0.193	 \\
$^5$A$_1$ &		0.94	&		0.06	&		0.02	&		1.00	&		0.06	&		0.94	&		0.140	 \\
$^5$B$_1$ &		0.01	&		0.99	&		0.02	&		1.00	&		0.01	&		0.99	&		0.215	 \\
$^5$B$_2$ &		0.44	&		0.96	&		0.57	&		0.43	&		0.04	&		0.56	&		0.088   \\
$^5$A$_2$ &		0.47	&		1.00	&		0.53	&		0.53	&		0.00	&		0.47	&		0.000	 \\
$^7$A$_1$ &		0.42	&		0.01	&		0.58	&		1.00	&		0.58	&		0.42	&		0.396	 \\
$^7$B$_1$ &		0.03	&		1.00	&		0.02	&		1.00	&		0.01	&		1.00	&		0.315	 \\
$^7$B$_2$ &		0.45	&		0.96	&		0.56	&		0.44	&		0.04	&		0.55	&		0.188	 \\
$^7$A$_2$ &		0.94	&		1.00	&		0.06	&		0.01	&		0.06	&		0.94	&		0.300   \\

\end{tabular}
\end{table}

\clearpage

\newpage

\section{Analytic Model}
We now describe an analytic model which allows us to directly join the cluster model with the bulk
thermodynamic properties of the solid.

The first step is to recognize the cluster's quartic-like inter-atomic potential
as two distinct harmonic potentials with a weak interaction (nearly diabatic).
Examining our cluster calculation for some reassurance that the minima are distinct, 
we approximate the zero point energies (ZPE) associated with the Sm-B$_2$ motion
using a quadratic fit to the ground state minima. 
The lower energy minimum has an estimated ZPE of about 3.4 ceV and the
higher energy minimum has an estimated ZPE of about 3.2 ceV. 
Both of these values are acceptable for considering the minima as distinct. 
(We take these estimates as higher than the true ZPEs as inclusion of 
anharmonic corrections will lower the values.)
Additionally, The high mass of the boron dimer ensures that the
harmonic oscillator wavefunction 
drop off rapidly away from equilibrium, so
we can neglect any boron tunneling between the two ground states.
It's important to consider the two wells independently because only then can we 
discuss a distinct oxidation state.

We continue with our diabatic-like approximation and assume that the timescale of switching between wells is much slower than the 
timescale for thermalization within a single well because of the significant jump of $\text{B}_2$
length between the two minima.

Because of this timescale separation, each harmonic oscillator has its own well-defined average energy given by the standard relation:  
\begin{equation}
\langle E \rangle = \hbar \omega_{\pm} \left( 1/2 + \left(\exp\left(\hbar \omega_{\pm} \beta\right) -1 \right)^{-1}\right) \pm \epsilon/2 
\label{eq:harmonic_oscillator_e}
\end{equation}
Here, $\epsilon$ is the difference between the minima, $\beta$ is the inverse temperature, and $\omega_{\pm}$ is the frequency of the upper / lower oscillator.

Considering transitions between wells we can treat the whole system as a two state system where each state has an energy  
given by equation \ref{eq:harmonic_oscillator_e}.
We are mostly interested in the energy difference between the two wells and
because the two frequencies are very similar, this difference is approximately given by $\epsilon$.

Considering the bulk solid as an ensemble of these two state clusters, 
states allows us to write temperature dependent expression for the average valence.
\begin{equation}
    V(T) = \frac{2.5 d + 3 \exp\left(-\beta\epsilon\right)}{d + \exp\left(-\beta \epsilon \right) }
\label{eq:valence}
\end{equation}
This was only possible because we are working with states that have a unique 
Sm valence state. Were we to work with adiabats instead, this simple expression would not be possible. 
We use the standard $\beta = \left(\text{k}_{\text{b}}T\right)^{-1}$,
and we set \textit{d} to the degeneracy of the lower level allowing it to vary in our expression rather than setting it to four like in figure 2A. This serves a validating purpose; it's a parameter that ought to be nearly four when fit with experimental valency data. 
We similarly expect the experimentally fit $\epsilon$ to agree with the cluster's energy difference.

One byproduct of our simplified model is that while we know that the motion of a dimer causes reduction of oxidation of the metal, 
the uncorrelated two state system does not provide insight into interactions between multiple clusters. 
For instance, this approach does not account for interactions between
sites. 
Rather than conjecturing here, we instead adopt, consistent with experimental data, a Sm average valence starting point of +2.5.

It is straightforward to calculate the excess specific heat provided by the two-level system.
Similarly, $C_{v}T^{-1}$ is easily derivable and we arrive at a slight variant on the textbook expression. 
\begin{equation}
\frac{d S}{d T} = \frac{1}{T^3}\frac{d \epsilon^2 \exp(\epsilon \beta)}{k_b\left(1 + d \exp\left(\epsilon \beta\right)\right)^2}    
\label{eq:specifcheat}
\end{equation}

In its current form, equation \ref{eq:specifcheat} is not directly comparable with experiment because there are 
a host of phonon modes in the solid which contribute to the specific heat. 
The well known specific heat profile of LaB$_6$ (a solid with a nearly identical crystal structure) 
allows us to extract all of the contributions we are not interested in. 
The only difference that we need to correct for doing this is the slight mass difference between Sm and La. 
Scaling equation \ref{eq:specifcheat} by some degeneracy ($\alpha$), adding the mass-scaled ($\beta$) specific heat of LaB$_6$, and adding a zero-temperature constant ($\gamma$) generates an equation that we can fit to experimental data.
\begin{equation}
    \frac{dS}{dT}\left(\text{SmB}_6\right) = \alpha\frac{dS}{dT}\left(\epsilon\right) + \beta \frac{dS}{dT}\left(\text{LaB}_6\right) + \gamma
\end{equation}

As seen in main text Figure 4D, the optimized parameters have clear physical meanings. $\alpha$ is the degeneracy of the total cluster system’s ground state surface, and $\beta$ is roughly the scaling of masses between Sm and La. 

Finally, to establish a connection between epsilon and the pressure we take epsilon to be a function of the average valence and manipulate equation \ref{eq:valence} to write an expression for the change in $\epsilon \text{k}_{\text{b}}^{-1}T^{-1}$.
\begin{equation}
    \frac{\epsilon}{\text{k}_{\text{b}}T} = - \ln\left(\frac{V d - d (2.5)}{3 - V}\right)
\end{equation}

\section{Phenomenological Band Model}
In the main text we briefly discussed the possibility that this model has the
components to explain the interesting field effects (namely, quantum oscillations) 
experimentally observed in SmB$_6$.
Quantitative theoretical exploration of the quantum oscillations is suitable for 
further study in its own right, so we will not
delve too deeply into this. 
However, here we present justification for that claim and propose a
microscopic Hamiltonian consistant with our model of the solid.

We start with a general overview of how quantum oscillations
can appear to begin with.
In a ``normal" system, it should not be possible to generate non-monotonic magnetization as a function of an applied field without violating thermodynamics -- magnetization measures the net alignment along the applied field, and it takes energy to move a magnetic spin away from the preferred field direction. 
This energy increases with the field, so the net alignment can never go down as the field increases.

The picture we have proposed for SmB$_6$ is not a ``normal'' situation, 
and the key feature of our model is that there are two energy minima
spaced within an energy such that a magnetic field of the order of Teslas can reverse the ordering of their energies. 
Additionally, every minimum has a different magnetic ordering and g factor. 
Within each minima the typical responses to applied fields are preserved 
(i.e. states with magnetism aligned with field decrease in energy with bigger fields),
but the additional degree of freedom allowing each Sm to switch between minima,
results in non-monotonicity of the magnetization.

This last statement can be derived by recognizing that magnetization is proportional to the 
derivative of Helmholtz free energy $(F = U - TS)$ with respect to the applied magnetic field.

\begin{equation}
    M = -\left(\frac{\partial F}{\partial B}\right)_T
\end{equation}
The key ingredient to having oscillations in $M$ is for the energy splittings not be strictly linear in
the applied magnetic field 
(cf. what happens in atoms in the presence of the spin-orbit
interaction). 
This readily happens in this model as the applied field induces level crossings 
and, critically, avoided level crossings between the different allowed states 
as a function of magnetic field.

We have thus far laid out that some form of coupling between the states must 
exist to produce the effects seen experimentally. 
To go further into this phenomenology it is necessary to propose
a microscopic Hamiltonian consistent with our model.
This Hamiltonian should be composed of two spin
lattices (one for each minima) coupled to each other based on their dynamical
occupations (i.e. each site must be in one minima or the other and not both). 
For the quantum oscillations measurements, an external magnetic field is added
in the standard way. 
Because the higher energy ($+\epsilon$) and lower energy ($0$)
minima are predominantly $^4A_2$ and $^6A_2$ respectively, we consider one lattice to be made of spin 5/2 particles and the other to be made of spin 3/2 particles. 
Each site is coupled to its neighbors by an exchange term $J_{i j}$.
Because only one spin exists on a site at a time, there is no coupling between spins on the same site.

These observations result in the following Hamiltonian for SmB$_6$:
\begin{equation}
\begin{aligned}
    \hat{H} = & 
    \epsilon \sum_{i} \hat{n}_{i}
    -\sum_{i}\vec{B} \cdot \left(\hat{n}_{i} \vec{S}^{(3/2)}_{i}
    + (1 - \hat{n}_{i})\vec{S}^{(5/2)}_{i}\right)\\
    -& \sum_{\langle i, j \rangle} J_{i j}
    \left(
    \hat{n}_{i} \vec{S}_{i}^{(3/2)} + (1-\hat{n}_{i})\vec{S}_{i}^{(5/2)}
    \right)
    \cdot
    \left(
    \hat{n}_{j}\vec{S}_{j}^{(3/2)} + (1-\hat{n}_{j})\vec{S}_{j}^{(5/2)}
    \right)
\end{aligned}
\end{equation}
Here, we denote the spin operators for these two particles on a specific spatial site $i$ as $\vec{S}^{(3/2)}_{i}$ and $\vec{S}^{(5/2)}_{i}$.
We also use the fermionic number operator $\hat{n}_{i}$ to signify the number of particles (either 0 or 1) in the s=3/2 state of $i$th site. 
We additionally have the constraint that each 
site must have a particle in either in the s=3/2 lattice
or the s=5/2 lattice.
Using the expression $(1-\hat{n}_i)$ in place of a number operator for the second
lattice trivially enforces this constraint.

This Hamiltonian is highly non-trivial to solve (even in the absence of an applied
field), containing a dynamical exchange interaction.
It is a worthy theoretical challenge on its own.
Nonetheless, we can extract some reassurance of it's physical description of SmB$_6$ 
by inspection of its form.
First, it has all the right pieces to allow for oscillations in the magnetization because of the spin-spin coupling between adjacent sites.
Second, it also has the potential to predict a mixed-valent ground state when the exchange interaction $J$ is negative and more destabilizing than the energy splitting of the clusters. 

\section{Discussion on the relationship of the $\mathrm{\textbf{SmB}}_{\mathbf{6}}$ model to other rare-earth hexaborides and dodecaborides}

How and why could the discovered vibronic structure of SmB$_6$ relate to that of other rare-earth hexaborides? They all have the same cubic structures, but dramatically varying properties.\cite{s5sakai2007effect,s6dorenbos2003systematic,7kim2014topological,s8goodrich1998fermi,s9weill1979electrical,s10fujita1980specific,s11hidaka2009specific,s12nozaki1980magnetic,s13kunii1985electronic,s14grechnev2012pressure,s15mccarthy1980magnetic,s16goodrich2006fermi,s17lazukov2016thermal,s18kuromaru2002multipolar,s19sera2004crystal,s20deng2013plutonium,s21xu2014direct,s22zhou2015pressure,s23akang2016electronic}
For example, EuB$_6$ and SmB$_6$ both have an anomalous peak in the specific heat, while for LaB$_6$ no such peak exists. T and p-sensitive magnetization, resistivity, and electronic phase transitions have been seen also in EuB$_6$, GdB$_6$, PrB$_6$, and YbB$_6$. GdB$_6$, and PrB$_6$ have easily accessible structural distortions. These observations resonate with different structural and electronic aspects of the SmB$_6$ model. It is likely that the types and strengths of the B-B and M-B bonds that these hexaborides can afford are the unifying factors governing their similarities and differences. For example, the radius of the rare-earth atom influences the lattice size, and thus the B-B separation in the dimers and the strength of the $\sigma(p_z)$-bond. The number of available \textit{f}-electrons in the rare-earths may play a role in strengthening or weakening the interaction with the boron (Ref 25, main text). Both effects would impact the accessibility of the possible bonded M-B/B-B states, the energy splitting between them, and the vibronic couplings, which could be the building blocks of a unifying model for the entire hexaboride series. Similar phenomenology has also recently been reported in YbB$_{12}$.\cite{s24abxiang2018quantum,s24balekseev2015high} Here, too, a similar kind of mixed valency is likely in play,\cite{s25kanai2015evidence} but with a more complex fundamental unit than a B$_2$ dimer, beyond the reach of the present computational tools.

\section{Raman spectroscopy}

\subsection{Phonon line shapes}
In the absence of electron-phonon coupling and disorder a line shape of a phonon is typically described by a Lorentz function, while disorder in a system leads to a change of phonon line shapes to Gaussian function.\cite{s26HollasSpectro} The width (typically defined as width at half maximum) of an observed phonon with a frequency ω is determined by disorder, if it is present, and a thermal population of the phonon levels with $\omega_{ph}=\omega/2$ through scattering on which the non-radiational decay of the excited phonons states occurs.\cite{s27kim2012temperature}
It follows the general formula 
$\Gamma\left(T,\omega\right) =  \Gamma_D + A\left( 2n_B \left(\frac{\omega}{2}\right) +1 \right)$
, where $\Gamma_D$ is a temperature independent term defined by disorder.

The coupling of the A$_{1\text{g}}$ phonon of SmB$_6$ to a continuum of interband electronic excitations results in a so-called Fano line shape of the phonon. This asymmetric line shape is a result of an interaction of a phonon mode with a background continuum, and can be described by an empirical formula 
$F\left(\omega,\omega_F,\Gamma_F,q\right) = \frac{1}{\Gamma_F q^2}\frac{\left[q + \alpha(\omega) \right]^2}{1 + \alpha(\omega)^2}$
, where $\alpha(\omega) = \frac{\omega - \omega_F}{\Gamma_F}$, $q$ is an empirical coupling parameter between the phonon and the electronic background.\cite{s28zhang2015observation,s29fano1961effects}

\subsection{Intensities of the features around 20meV}
\begin{figure}[h!]
   \includegraphics[width=0.45\linewidth]{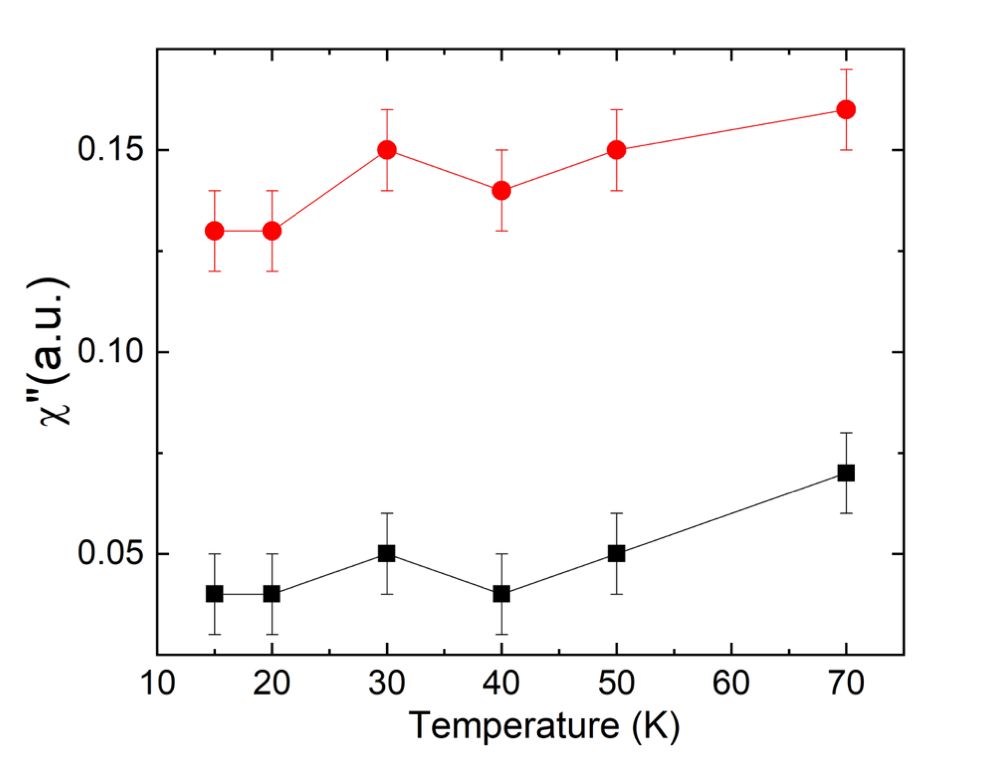}
   \caption{Intensities of the features which appear due to electron-phonon coupling as a result of valence fluctuations in the region of 20 meV. Intensity of the narrow phonon feature is shown in red, intensity of the exciton-polaron feature is shown in black.}
\end{figure}

\subsection{Experimental Details}
Raman scattering was performed for freshly cleaved surfaces of single crystals of SmB$_6$. Orientation of measured crystals and surfaces was based on XRD and confirmed by measurements of polarized Raman scattering. Low temperature measurements were performed using cold finger cryostat. To prevent heating of the sample, laser power was kept below 10 mW for a laser spot of approximately 100$\mu$m $\times$ 50$\mu$m. 

\bibliographystyle{naturemag}
\bibliography{si_citations}